\documentclass[12pt]{article}
\usepackage{epsfig, epsf, graphicx}
\usepackage{pstricks, pst-node, psfrag}
\usepackage{amssymb,amsmath,bm}
\usepackage{verbatim,enumerate}
\usepackage{rotating, lscape,natbib}
\usepackage{setspace}
\usepackage{multirow}
\usepackage{tikz}
\usetikzlibrary{arrows,calc,tikzmark}
\usepackage{hyperref}
\usepackage{float}
\usepackage{caption}
\usepackage{subcaption}
\usepackage{animate}
\usepackage{multimedia}
\usepackage{appendix}
\usepackage{comment}
\usepackage{url}
\usepackage[utf8]{inputenc}
\usepackage{bbm}
\usepackage[english]{babel}
\usepackage{xcolor}
\usepackage{natbib}

\usepackage{color}
\usepackage{verbatim}

\newcommand{\bs}{\boldsymbol}
\newcommand{\mb}{\mathbf}

\begin{document}

\thispagestyle{empty} \baselineskip=28pt \vskip 5mm
\begin{center} {\Huge{\bf Improving Bayesian Local Spatial Models in Large Data Sets}}
\end{center}

\baselineskip=12pt \vskip 10mm

\begin{center}\large

Amanda Lenzi\footnote[1]{
\baselineskip=10pt Statistics Program,
King Abdullah University of Science and Technology,
Thuwal 23955-6900, Saudi Arabia.}, 
Stefano Castruccio\footnote[2]{
\baselineskip=10pt Department of Applied and Computational Mathematics and Statistics,
University of Notre Dame, Notre Dame, IN 46556, USA.\\
This publication is based on research supported by the King Abdullah University of Science and Technology (KAUST) Office of Sponsored Research (OSR) under Award No: OSR-2018-CRG7-3742.},
H\aa vard Rue$^{1}$,
and Marc G. Genton$^{1}$
\end{center}

\baselineskip=17pt \vskip 10mm \centerline{\today} \vskip 15mm

\begin{center}
{\large{\bf Abstract}}
\end{center}

Environmental processes resolved at a sufficiently small scale in space and time inevitably display non-stationary behavior. Such processes are both challenging to model and computationally expensive when the data size is large. Instead of modeling the global non-stationarity explicitly, local models can be applied to disjoint regions of the domain. 
The choice of the size of these regions is dictated by a bias-variance trade-off; large regions will have smaller variance and larger bias, whereas small regions will have higher variance and smaller bias. 
From both the modeling and computational point of view, small regions are preferable to better accommodate the non-stationarity. However, in practice, large regions are necessary to control the variance. We propose a novel Bayesian three-step approach that allows for smaller regions without compromising the increase of the variance that would follow. We are able to propagate the uncertainty from one step to the next without issues caused by reusing the data. The improvement in inference also results in improved prediction, as our simulated example shows. We illustrate this new approach on a data set of simulated high-resolution wind speed data over Saudi Arabia.

\baselineskip=14pt

\par\vfill\noindent
\par\medskip\noindent
{\bf Short title}: Local Spatial Models

\clearpage\pagebreak\newpage \pagenumbering{arabic}
\baselineskip=26pt

\section{Introduction}\label{sec:intro}

The rising popularity of statistical methods for environmental data calls for the development of new methods that are able to capture the underlying varying dependencies and to provide computationally efficient inference for the ever increasing amount of data. Traditional geostatistical approaches are not only computationally intensive but are also based on stationarity assumptions, which is convenient but too restrictive and rarely realistic. For instance, wind at sufficiently small temporal resolution (e.g., hourly or sub-hourly) tends to be more variable over complex terrain than over flat surfaces due to geographical features creating eddies. Additionally, failing to account for how physical processes such as weather patterns vary over time or space can lead to an unrealistic assessment of the dependence, and hence suboptimal inference and prediction.

Traditionally, methods have focused on characterizing the spatial and spatio-temporal non-stationarity explicitly via the covariance function. The deformation method in \citet{sampson1992nonparametric} constructs a non-stationary covariance structure from a stationary structure by re-scaling the spatial distance, which was subsequently extended to the Bayesian context in \citet{damian2001bayesian} and \citet{schmidt2003bayesian2}. Another class of non-stationary methods is built on the process convolution or kernel smoothing method, introduced by \citet{higdon1998process}, which uses a spatially varying kernel and a white noise process to create the covariance structure. Other well-known approaches to model non-stationarity include representing the covariance function as a linear combination of basis functions and modelling the covariance matrix of the random coefficients \citep{nychka2002multiresolution}, and to account for the effect of covariate information directly in the covariance function \citep{schmidt2011considering, neto2014accounting}. For a review on the existing literature on non-stationary methods, see \citet{risser2016nonstationary}.

Although all of the above methods produce valid models, their computational burden for inference and prediction can be unfeasible for large data sets. 
Indeed, for evaluating a Gaussian likelihood in a data set of size $n$, $O(n^2)$ entries need to be stored and $O(n^3)$ flops need to be computed for the log-determinant and matrix factorization. This task is feasible in modern computers only when $n$ is at most a few tens of thousands of points. Additionally, evaluating a non-stationary model implies inference on a larger parameter space, which requires an exponentially increasing number of likelihood evaluations for frequentist inference or posterior sampling \citep{edw19}. To address the difficulties in computation for large data sets, \citet{nychka2018modeling} used a multi-resolution representation of Gaussian processes to represent non-stationarity based on windowed estimates of the covariance function under the assumption of local stationarity, and successfully used this idea to emulate fields from climate models. 
\citet{kuusela2018locally} proposed modelling Argo profiling float data using locally stationary Gaussian process regression, where parameter estimation and prediction were carried out in a moving window. Other works related to moving window methods have been developed and applied in \citet{hammerling2012mapping} and \citet{tadic2015mapping} to model remote sensing data. 

The seminal work of \citet{lindgren2011explicit} predicated avoiding modeling the covariance function altogether and modeled the data via a Stochastic Partial Differential Equation (SPDE) instead. By considering a spatial field as a solution of an SPDE, and describing the covariance function only implicitly, inference is of the order $O(n^{3/2})$ \citep{rue2017bayesian}, thus allowing inference on considerably larger data sets than covariance-based methods. The computational benefits arise from the precision matrix (inverse covariance matrix) resulting from the approximate stochastic weak solutions of the SPDE, which has a Markovian structure where only close neighbours are non-zero \citep{rue2005gaussian}. By spatially varying the coefficients in the SPDEs, it is also possible to construct a variety of non-stationary models. \citet{bolin2011spatial} developed such a method for global ozone mapping, whereas \citet{bakka2019non} defined a continuous solution to an SPDE with spatially varying coefficients for solving problems that involve a physical barrier to spatial correlation. By combining the SPDE representation of a stationary Mat\'{e}rn field with the deformation method, \citet{hildeman2019spatial} modelled non-stationarity in significant wave heights. Locally non-stationary fields were considered in \citet{fuglstad2015exploring} by letting the coefficients in the SPDE vary with position, and further discussed and generalized for spatially varying marginal standard deviations and correlation structure in \citet{fuglstad2015does}. More recently, \cite{gei20} formulated a global SPDE model with locally varying coefficients with a change of structure across land and ocean. Another application of the SPDE approach to model non-stationarity is to include covariates directly into the model parameters; see \citet{ingebrigtsen2014spatial} for an application to annual precipitation in Norway. 

\begin{figure}[htb!]
	\centering
	\begin{subfigure}[b]{0.53\textwidth}
		\includegraphics[width=\textwidth,height=7.5cm]{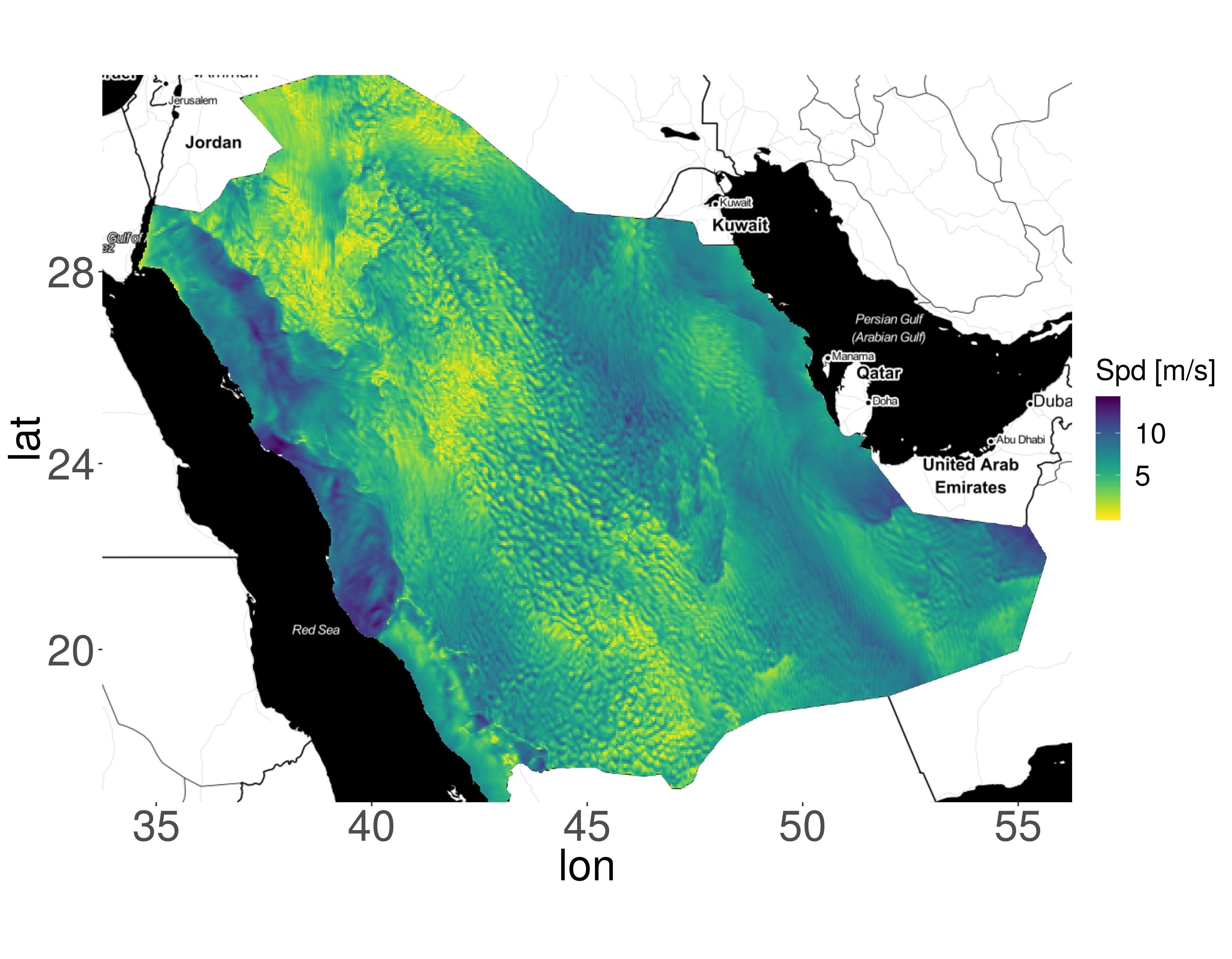}
			\caption{} 
			\label{fig:map}
	\end{subfigure} 
	\begin{subfigure}[b]{0.45\textwidth}
		\includegraphics[width=\textwidth,height=6.7cm]{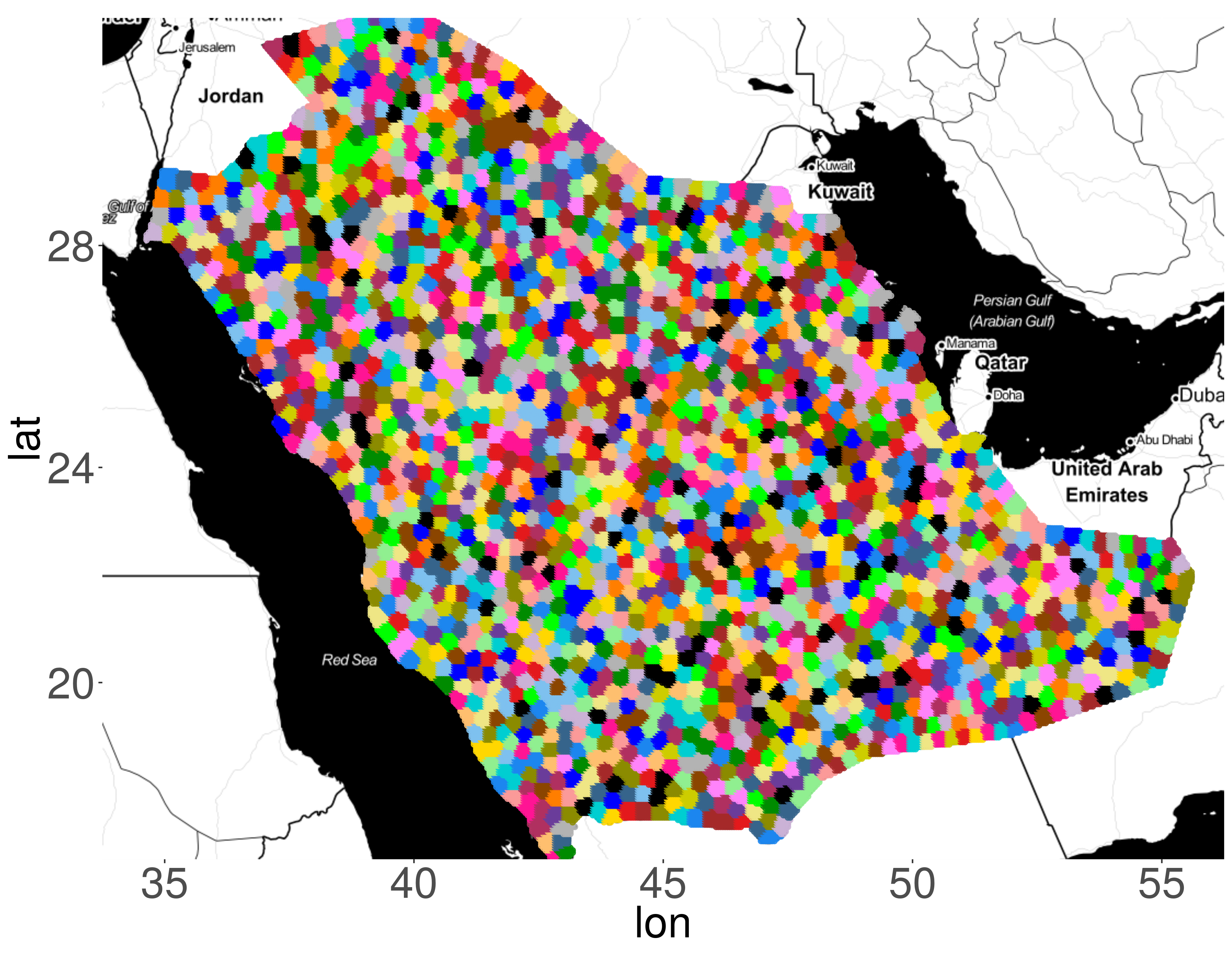}
			\caption{} 
			\label{fig:cl}
	\end{subfigure}
		\caption{(a): Snapshot of 2 meters wind speed simulations at 84,494 locations over Saudi Arabia on 03/06/2010 averaged between 14:00 and 15:00 local time.
		The minimum wind speed is around 0.3 m/s and the maximum is 14 m/s.
		(b): Location of the $R=2000$ clusters.}
		\label{fig:map-data-cl}
\end{figure}

The aim of this paper is to develop a new method for modelling large data sets with spatial dependence that not only improves local models in terms of inference and prediction, but is also computationally affordable. As a motivating example, we use the high-resolution simulated wind data from a computer model displayed in Figure~\ref{fig:map}. We partition this data into several small disjoint subsets of the data, which we call ‘regions’, as shown in Figure~\ref{fig:cl}. Modeling and predicting such variable over a large region present several challenges. First, the data structure at this high resolution is very complex, with details and features that are difficult to capture with a single model. As a consequence, the assumption of stationarity for the entire region is inappropriate. Second, because of the large number of locations, we need a method that is computationally efficient. We show that our method is able to address not only the modeling challenges arising from the inherent non-stationarity of hourly wind, but also the computational issues that are implied by the large data size.

When choosing the size of these regions, we face the conflicting issue of bias-variance trade-off in parameter estimation. Ideally, one wants to choose regions that accurately capture the features in the data (low variance), but also have high predictive out-of-sample skills (low bias). Indeed, small regions reduce the model bias and allow fast computations, at the expense of low accuracy (high variance) in the parameter estimation. Large regions instead allow a control of the variance but also imply a sub-optimal characterization of the dependence structure, hence a bias. 

We propose a novel three-step approach, which simultaneously allows for small regions and low variance. The key is to allow small regions to model the local dependence, and correct the estimated parameter distribution with a smoothing step that borrows strength from neighboring regions. 
The smoothing step is performed so that it accounts for the uncertainty  of the parameter estimates from the first step. 
The resulting smoothed distribution represents the adjusted uncertainty of the local parameters, which is then used for re-fitting the models. Allowing this adjusted uncertainty to be used as a new prior would imply the incorrect premise of the model being influenced by the data twice, hence our approach restricts the information propagation by including it as the new posterior estimates instead. Crucially, the approach we propose is  computationally fast and scalable to massive spatial data sets, as it can be fully parallelized across regions. We start with a simple example where the new posterior is the mode of the distribution from the smoothing step. 
Then, using the wind data in Figure~\ref{fig:map}, we show that it is possible improve the predictive performances by also allowing the uncertainty to propagate from one step to the next.

\begin{figure}[htb!]
	\centering
	\begin{subfigure}[b]{0.49\textwidth}
		\includegraphics[width=1\textwidth]{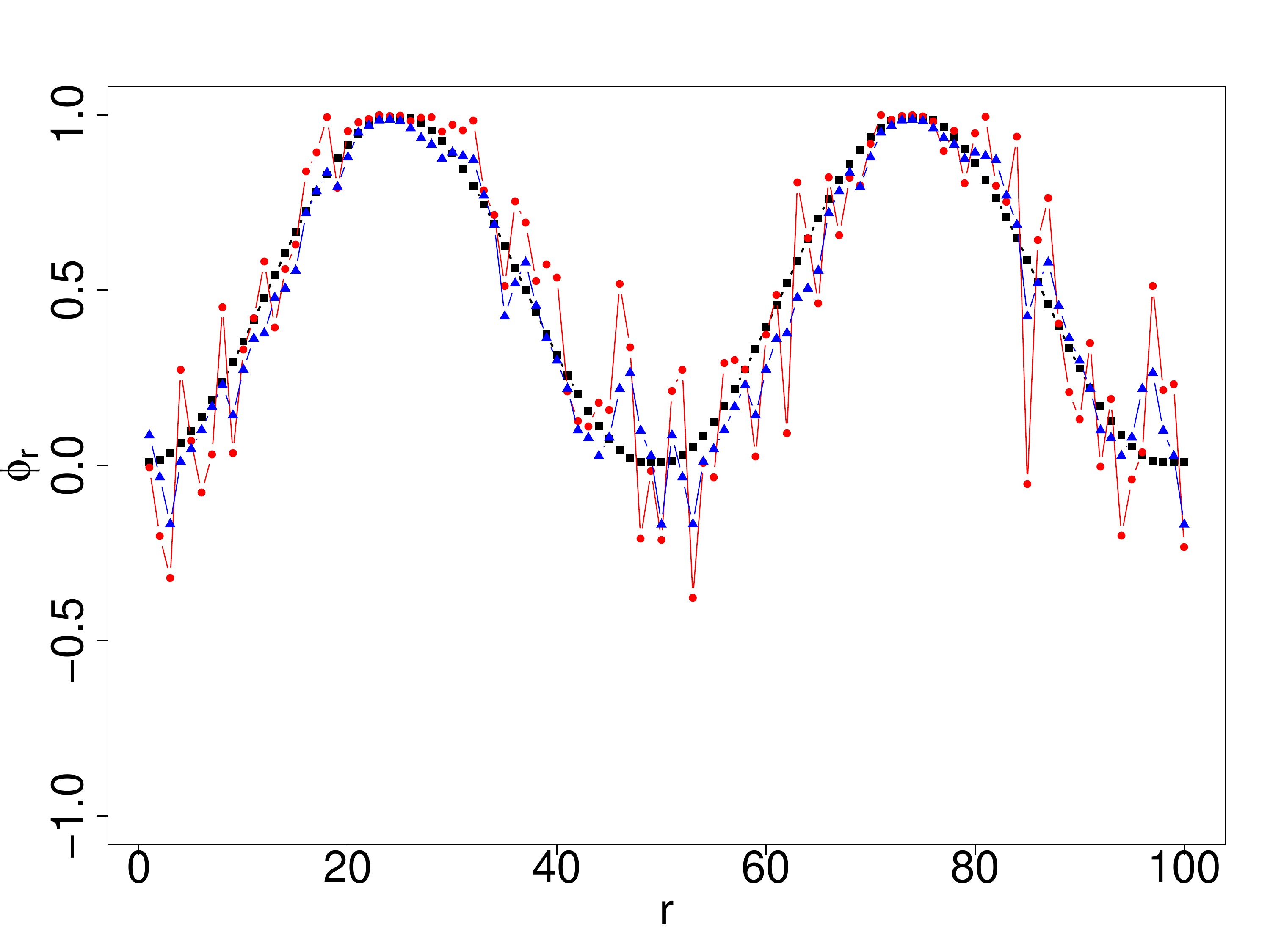}
			\caption{} 
			\label{fig:tau2}
	\end{subfigure}	
	\begin{subfigure}[b]{0.49\textwidth}
		\includegraphics[width=1\textwidth]{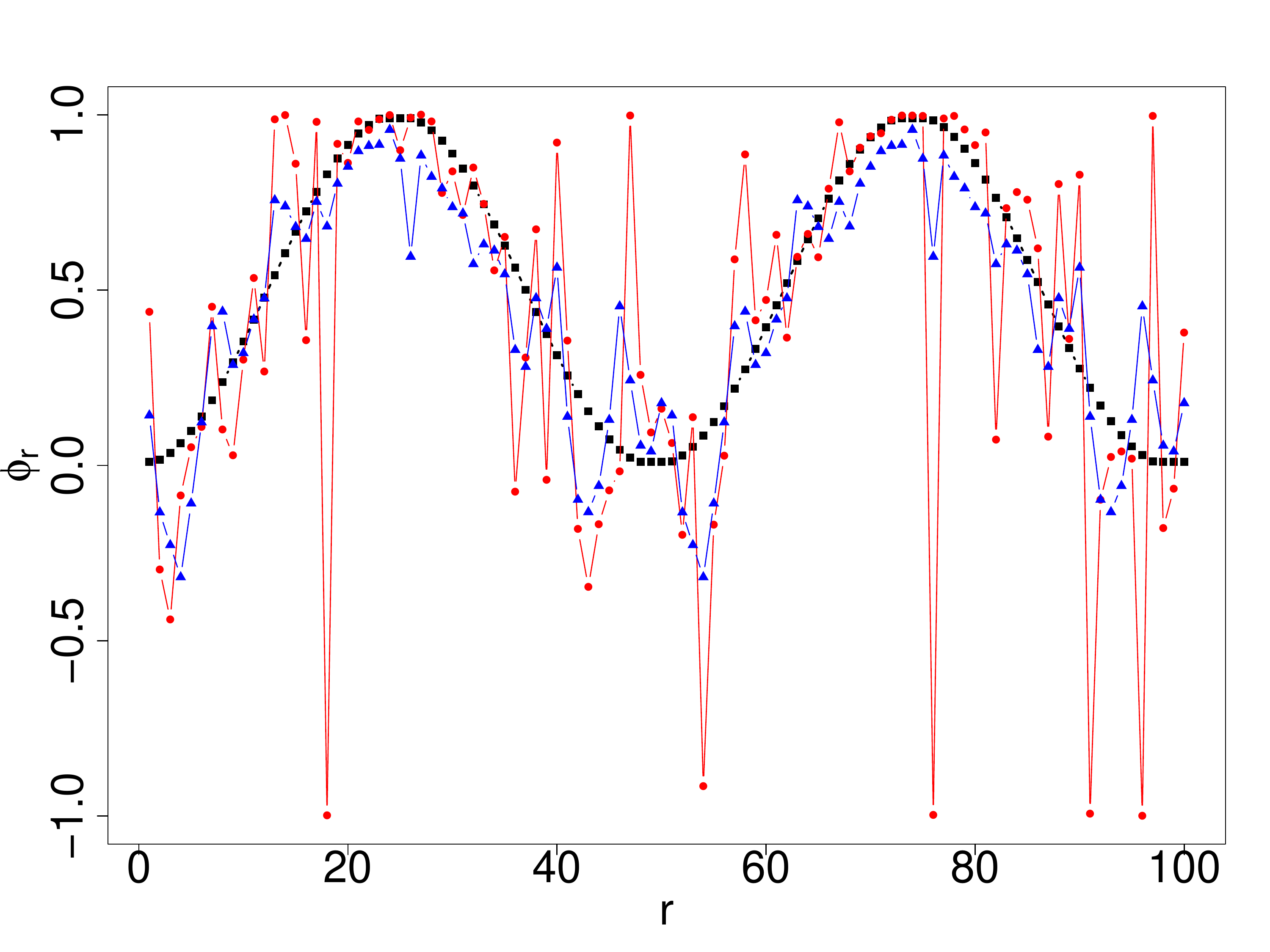}
	\caption{} 
	\label{fig:tau1}
	\end{subfigure} 
	\caption{Values of $\phi_{r}$ used to simulate $R=100$ time series from (\ref{eq:ar1-ex}) of length 50 (black squares), estimated values of $\phi_{r}$ from fitting the AR(1) model to the simulated data (red circles), and estimates after fitting a smoothing spline (blue triangles). The left plot corresponds to simulations with fixed  $\tau=2$ and the right plot corresponds to $\tau=1$.}
\label{ar1-coefs}
\end{figure}

Our three-step approach is best exemplified by considering a toy data set, where each region consists of an autoregressive process of order one, AR(1). We simulate $R$ time series from this model, where each time series contains $T$ observations, $\mb{y}_r = \{y_{r}(1), \ldots, y_{r}(T)\}^{\top}$. For each $r$, the observations $\mb{y}_r$ are assumed to be conditionally independent, given the latent Gaussian random field $\mb{x}_r = \{x_{r}(1), \ldots, x_{r}(T)\}^{\top}$ and the hyperparameter $\phi_r$: 
\begin{equation}
\begin{aligned} 
y_{r}(t) & = x_{r}(t) + \epsilon_{r}(t),  \quad \epsilon_{r}(t) \overset{\text{iid}} \sim \mathcal{N}(0, 1/\tau), \\
x_{r}(t) & = \phi_{r} x_{r}(t-1) + \omega_{r}(t), \quad  \omega_{r}(t) \overset{\text{iid}} \sim \mathcal{N}(0, 1),
\end{aligned}
\label{eq:ar1-ex}
\end{equation}
where $t = 2, \ldots, T$ is an index for time, $|\phi_{r}| < 1$  and $\tau$ is the fixed precision (known and the same for all time series).
Figure~\ref{ar1-coefs} shows the different values of $\phi_{r}$ used to simulate $R=100$ time series from \eqref{eq:ar1-ex}, where $\phi_{r}$ changes according to a series of sine squared (black squares in Figure~\ref{ar1-coefs}). For each time series, we set $T=50$ and two different values for the precision: $\tau=2$ and $\tau=1$ in Figure~\ref{fig:tau2} and \ref{fig:tau1}, respectively.
In the first step, we estimate local models for each time series (red circles in Figure~\ref{ar1-coefs}). 
In the second step, we apply a correction on the parameters' estimates from the first step, based on information from neighbouring regions (blue triangles in Figure~\ref{ar1-coefs}). 
The third step consists of re-fitting the model in \eqref{eq:ar1-ex} to each time series, propagating the information from the adjusted posterior estimates from the second step back into the analysis.
Figure~\ref{ar1-coefs} shows that our correction improves the parameter estimates substantially not only for the more extreme case where $\tau=1$ in panel (b), but also when $\tau=2$ in panel (a). More details on this example will be provided in Section~\ref{sec:ar1}.

The remainder of this paper is organized as follows. In Section~\ref{sec:method} we provide an overview of the proposed methodology. Further details of our approach using the AR(1) example are given in Section~\ref{sec:ar1}. The application to the wind speed data in Figure \ref{fig:map-data-cl} is presented in Section~\ref{sec:wrf}. A comprehensive discussion and conclusions are provided in Section~\ref{sec:disc}. 
 
\section{Overview of the proposed methodology} \label{sec:method}

\subsection{Background} \label{sec:background}

We consider a non-stationary and possibly very large data set, and a partition of the domain into regions where the assumption of stationarity is plausible, defined as $\Omega_r , r = 1,\ldots, R$, where each observation is associated with exactly one $\Omega_r$. Each region contains $N_r$ observations, $\mb{y}_r = \{y_{r}(1), \ldots, y_{r}(N_r)\}^{\top}$. For each $\Omega_r$, consider the following hierarchical structure:
\begin{equation}
    \begin{aligned}
    \mb{y}_r \mid \mb{x}_r ,\bs{\theta}_{r}  & \sim \prod_{i=1}^{N_r} \pi \{ y_{r}(i) \mid x_{r}(i), \bs{\theta}_{r} \}, \\
    \mb{x}_r \mid \boldsymbol{\theta}_{r}  & \sim \pi(\mb{x}_r \mid \boldsymbol{\theta}_{r}), \\
    \boldsymbol{\theta}_{r}  & \sim \pi(\boldsymbol{\theta}_{r}),
    \end{aligned}
    \label{eq:hierarq}
\end{equation}
where $\mb{x}_r = \{x_{r}(1), \ldots, x_{r}(N_r)\}^{\top}$ is the vector of the latent field that describes the underlying spatial dependence structure, $\boldsymbol{\theta}_{r}$ is the $m$-dimensional vector of hyperparameters and $\pi$ is a generic distribution. The observations $\mb{y}_r$ are assumed to be conditionally independent, given $\mb{x}_r$ and $\bs{\theta}_{r}$. The resulting joint posterior distribution of $\mb{x}_r$ and $\bs{\theta}_{r}$ is given by
\[
\pi(\mb{x}_r, \bs{\theta}_{r} \mid \mb{y}_r) \propto \pi(\bs{\theta}_{r}) \pi(\mb{x}_r \mid  \bs{\theta}_{r}) \prod_{i =1}^{N_r} \pi\{y_{r}(i) \mid x_{r}(i), \bs{\theta}_{r}\}. 
\]
Our main goal is to extract the posterior marginal distributions for the elements of the latent field, $\pi\{x_r(i) \mid \mb{y}_r\}$ and hyperparameters, $\pi\{\theta_{r}(j) \mid \mb{y}_r\}$, and use them to obtain predictive distributions at unsampled locations. Calculation of these univariate posterior distributions requires integrating with respect to $\mb{x}_r$ and $\bs{\theta}_{r}$:
\begin{equation}
\begin{aligned}
\pi\{x_r(i) \mid \mb{y}_r\} & = \int \pi(x_r(i) \mid \mb{y}_r, \bs{\theta}_{r})  \pi(\bs{\theta}_{r} \mid \mb{y}_r) \mbox{d} \bs{\theta}_{r}, \quad i = 1, \ldots, N_r, \\
\pi\{\theta_{r}(j) \mid \mb{y}_r\} & = \int \pi(\bs{\theta}_{r} \mid \mb{y}_r) \mbox{d} \bs{\theta}_{r}(-j), \quad j = 1, \ldots, m,
\label{eq:margpost}
\end{aligned}
\end{equation}
where $\bs{\theta}_{r}(-j)$ is the vector of all but the $j$-th hyperparameter component omitted. When the integrals in \eqref{eq:margpost} cannot be found analytically, approximations are typically obtained via simulation-based methods such as MCMC.
Alternatively, \citet{rue2009approximate} proposed an approximate Bayesian inference approach that has become increasingly popular in the last decade. Approximations for $\pi(x_r(i) \mid \mb{y}_r, \bs{\theta}_{r})$ and  $\pi(\bs{\theta}_{r} \mid \mb{y}_r)$ are obtained via a Laplace approximation (see \citet{rue2017bayesian} for a comprehensive review on this approximation). The posterior $\pi(\bs{\theta}_{r} \mid \mb{y}_r)$ is computed as
\[
\pi(\bs{\theta}_{r} \mid \mb{y}_r)\approx \left.\frac{\pi(\mb{y} \mid \mb{x},\bs{\theta})\pi(\mb{x} \mid \bs{\theta})\pi(\bs{\theta})}{\tilde{\pi}(\mb{x} \mid \mb{y},\bs{\theta})}\right|_{x=x^*(\bs{\theta})}=\tilde{\pi}(\bs{\theta}_{r} \mid \mb{y}_r),
\]
where $\tilde{\pi}(\mb{x} \mid \mb{y},\bs{\theta})$ is a Laplace approximation, and $x^*(\bs{\theta})$ is the mode of $\mb{x}$ for a specific value of $\bs{\theta}$. Similarly we obtain $\tilde{\pi}(x_r(i) \mid \mb{y}_r, \bs{\theta}_{r})$, the approximation of $\pi(x_r(i) \mid \mb{y}_r, \bs{\theta}_{r})$.
These are then used to construct the following nested approximations

\begin{equation}
\begin{aligned}
\tilde{\pi}\{x_r(i) \mid \mb{y}_r\} & = \int \tilde{\pi}(x_r(i) \mid \mb{y}_r, \bs{\theta}_{r})
\tilde{\pi}(\bs{\theta}_{r} \mid \mb{y}_r) \mbox{d} \bs{\theta}_{r}, \quad i = 1, \ldots, N_r, \\
\tilde{\pi}\{\theta_{r}(j) \mid \mb{y}_r\} & = \int \tilde{\pi}(\bs{\theta}_{r} \mid \mb{y}_r) \mbox{d} \bs{\theta}_{r}(-j), \quad j = 1, \ldots, m.
\label{eq:margpostapp}
\end{aligned}
\end{equation}

\subsection{Improving the local estimates}

We propose a new method for improving the estimation of $\tilde{\pi}\{\bs{\theta}_{r} \mid \mb{y}_r\}$ in \eqref{eq:margpostapp} and hence also improving  the estimated $\tilde{\pi}\{x_r(i) \mid \mb{y}_r\}$, for $i = 1, \ldots, N_r$. Since each region is selected to be small enough to approximate the local non-stationarity well, the resulting parameters' estimates are likely to have a large variance, and smoothing across the regions is used to reduce it. 

The method is based on two extra steps in the estimation procedure from the previous section.
In Step 2 we apply a correction to the posteriors $\tilde{\pi}(\bs{\theta}_{r} \mid \mb{y}_r)$ by smoothing the mode of this distribution across $r$. In Section~\ref{sec:ar1}, we show a one dimensional  example with a smoothing spline, while in Section~\ref{sec:windresults} we describe the two dimensional case with a spatial model. We denote by $\tilde{\pi}_{\mbox{\scriptsize{smooth}}}(\bs{\theta}_{r} \mid \mb{y})$ the resulting smoothed distribution for region $r$ in Step 2 of our approach, where $\mb{y}$ is the combined data sets from all regions, i.e.,
$\mb{y} = (\mb{y}_{1}^{\top}, \ldots, \mb{y}_{R}^{\top})^{\top}$. 
In Step 3, the correction from Step 2 is propagated back into the analysis as the posterior for each region: 
\begin{equation}
\begin{aligned}
\tilde{\pi}_{\mbox{\scriptsize{smooth}}}\{x_r(i) \mid \mb{y}_r\} & = \int \tilde{\pi}(x_r(i) \mid \mb{y}_r, \bs{\theta}_{r})  
\tilde{\pi}_{\mbox{\scriptsize{smooth}}}(\bs{\theta}_{r} \mid \mb{y}) \mbox{d} \bs{\theta}_{r}, \quad i = 1, \ldots, N_r, \\
\tilde{\pi}_{\mbox{\scriptsize{smooth}}}\{\theta_{r}(j) \mid \mb{y}_r\} & = \int \tilde{\pi}_{\mbox{\scriptsize{smooth}}}(\bs{\theta}_{r} \mid \mb{y}) \mbox{d} \bs{\theta}_{r}(-j), \quad j = 1, \ldots, m,
\end{aligned}
\label{eq:approxx2}
\end{equation}
where $\tilde{\pi}({x}_r(i) \mid \mb{y}_r, \bs{\theta}_{r})$ is obtained by plugging values of $\bs{\theta}_{r}$ from $\tilde{\pi}_{\mbox{\scriptsize{smooth}}}(\bs{\theta}_{r} \mid \mb{y}_r)$ obtained in Step 2. Step 3 is very computationally efficient, since the posteriors for the hyperparameters have already been estimated, and as in Step 1 the models for each region can be fully parallelized.
Also, as the posterior marginals in \eqref{eq:approxx2} are the basis to derive the predictive distributions, the proposed correction will also have a direct impact in prediction performance.

Here, the vector $\bs{\theta}_r$ contains the hyperparameters that need to be smoothed, while the ones that do not require the smoothing are included in $\mb{x}_r$. In practice, it is more important to smooth hyperparameters that have a higher variability and are harder to estimate.

Our approach has a crucial difference compared to Empirical Bayes methods. The key is to account for the information from the smoothing in Step 2 directly into the posterior distribution in Step 3, as opposed to introducing it through priors as in Empirical Bayes methods. By doing so, we prevent the estimation in Step 3 to be influenced by the likelihood of the data that was already used in Step 1, and thus avoiding using the data twice. Moreover, our approach allows for uncertainty propagation from Step 2 to Step 3.

\section{Simulation with spatially varying AR(1) process} \label{sec:ar1}

\subsection{Model description}

In the Introduction we briefly introduced our method on a simulated example (see Figure~\ref{ar1-coefs}) using the AR(1) model in \eqref{eq:ar1-ex}. Here, we provide all the details about the methodology in light of the steps proposed in the previous section. For the ease of exposition, we fix the precision $\tau$ in \eqref{eq:ar1-ex}, so that for each region $r$ only the hyperparameter $\phi_r$ needs to be estimated. No covariates or additional random effects have been included in \eqref{eq:ar1-ex}, but the steps below can be easily adapted to account for them. 

The model is a special case of the hierarchical framework proposed in \eqref{eq:hierarq}. Indeed for the first equation of the hierarchy, the likelihood of the data $\mb{y}_r$ given the latent field $\mb{x}_r$ and the hyperparameter $\phi_r$ is given by
\[
    \mb{y}_r \mid \mb{x}_r, \phi_r \sim \mathcal{N}_{T}(\mb{x}_r,\tau^{-1} \textbf{I}_T),
\]
where $\textbf{I}_T$ is the  $T \times T$ identity matrix and $\tau$ is the fixed precision, while $\mathcal{N}_{T}$ is a $T$-dimensional normal distribution. For the latent process $\mb{x}_r$, we assume that the marginal distribution of $x_r(1)$ is Gaussian with mean zero and variance $1/(1-\phi_r^2)$ to have a stationary process. The joint distribution can be written as 
\[
\pi(\mb{x}_r \mid \phi_r) \sim \mathcal{N}_{T}(\mb{0},\mb{Q}^{-1}_{x,r}),
\]
where $\mb{Q}_{x,r}$ is the tridiagonal precision matrix of an AR(1) process. 

The three steps of our approach can be summarized as follows:

{\bf{Step 1: The model fitted to each region}}. 
Fit the AR(1) model in (\ref{eq:ar1-ex}) with fixed known $\tau$ to each time series, $\mb{y}_r$, separately. Following the notation in Section~\ref{sec:method}, we define the variance-stabilizing transformation $\bs{\theta}_r=\theta_r = \mbox{log} \Big( \frac{1+\phi_r}{1-\phi_r} \Big)$, where $\theta_r$ has a normal prior with mean zero and precision 0.15, independent across $r$. We then obtain the posterior marginal distributions for the latent field and for the hyperparameter $\theta_r$, which we denote by $\tilde{\pi}\{x_r(t) \mid \mb{y}_r\}$ and $\tilde{\pi}(\theta_r \mid \mb{y}_r)$, respectively, for $t=1, \ldots, T$ and $r=1, \ldots, R$. Inference is performed using the \textit{R-INLA} package \citep{rue2009approximate}.

{\bf{Step 2: Smoothing the hyper-parameter}}. As in \citet{lindgren2008second}, we assume a continuous spline on a discrete set of knots with a second order random walk RW(2).
We denote by ${\hat{\theta}_r}$ the mode for $\tilde{\pi}(\theta_r \mid \mb{y}_r)$ from Step 1, and we assume a normal distribution: ${\hat{\theta}_r} \sim \mathcal{N}(u_r, \tau_{\theta;r}^{-1})$, where $ \tau_{\theta;r}$ is the precision and is such that  $\mbox{log}(\tau_{\theta;r}) = \mbox{log}(1/\widehat{\mbox{sd}}_{r}^2)$, where $\widehat{\mbox{sd}}_{r}$ is the estimated standard deviation of the posterior distribution $\tilde{\pi}(\theta_r \mid \mb{y}_r)$. 
The vector ${\bs{u}} = ({u_1}, \ldots, {u_R})^{\top}$ is assumed to have independent second-order increments:
\begin{equation}
    \Delta^2 {u_r} = {u_r} - 2 {u_{r+1}} + {u_{r+2}} \sim \mathcal{N}(0, \tau_{u}^{-1}), \quad r=1,\ldots, R-2,
     \label{eq:rw2}
\end{equation}
where $\tau_{u}$ is the precision parameter and can be used to control the degree of smoothing across regions. Section~\ref{sec:smooth} discusses a method for choosing the optimal value of $\tau_{u}$. 

{\bf{Step 3: Re-fit the model to each region using the estimated mode}}. 
For each region $r$, we assume that the posterior distribution for the hyperparameters, namely $\tilde{\pi}_{\mbox{\scriptsize{smooth}}}(\theta_r \mid \mb{y}_r)$, is a point mass concentrated at the mode of $\hat{\theta}_r$ from Step 2. Our choice was dictated by ease of exposition, and in the wind data application in Section~\ref{sec:windresults} we will show a more general approach with integration points and weights instead of just the mode.
The marginal posterior for the latent process $\mb{x}_r$ is then obtained from the first equation in \eqref{eq:approxx2}.
Because here there are no hyperparameters that need to be re-estimated in this example, re-fitting the model is equivalent to updating the posterior for $\mb{x}_r$ given the data under the smoothing from Step 2.

Step 3 implies a change of the original posterior in Step 1, and hence a change in the prior of the model. While retrieving the appropriate prior is not relevant for our method, it is still however possible, and in the Appendix we show the steps to do so. Figure~\ref{fig_priors} shows (a) the log posterior distributions, (b) log likelihood function, and (c) log prior distributions from Step 1 (solid red) and Step 2 (dashed blue), for $\phi_r=0.88$. The log prior distributions were obtained simply by subtracting the log likelihood from the log posterior distributions, and the vertical line represents the true value. The proposed smoothing in Step 2 concentrates the posterior (and consequently the prior) considerably closer to the true value $\phi_r$ than a standard approach with no smoothing. Similar results can be observed for other choices of $\phi_r$; the mean absolute error across $r$ of the estimated mode posterior distributions from Steps 1 and 2 are $0.23$ and $0.08$, whereas the mean absolute error of the estimated mode priors for these steps are $0.61$ and $0.09$, respectively.

\begin{figure}[htb!]
	\centering
	\begin{subfigure}[b]{0.328\textwidth}
		\includegraphics[width=1\textwidth]{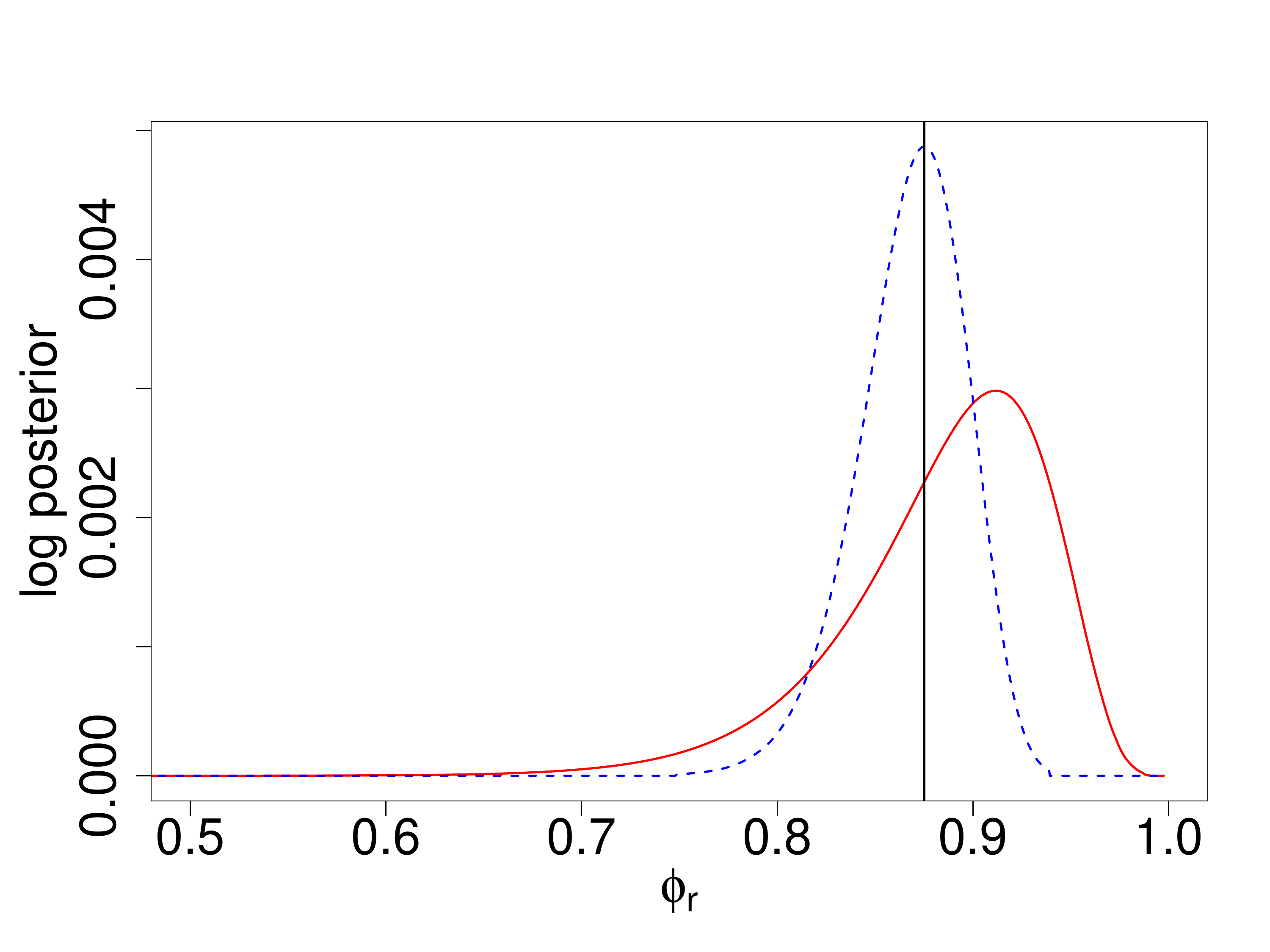}
		\caption{}
	\end{subfigure}	
	\begin{subfigure}[b]{0.328\textwidth}
		\includegraphics[width=1\textwidth]{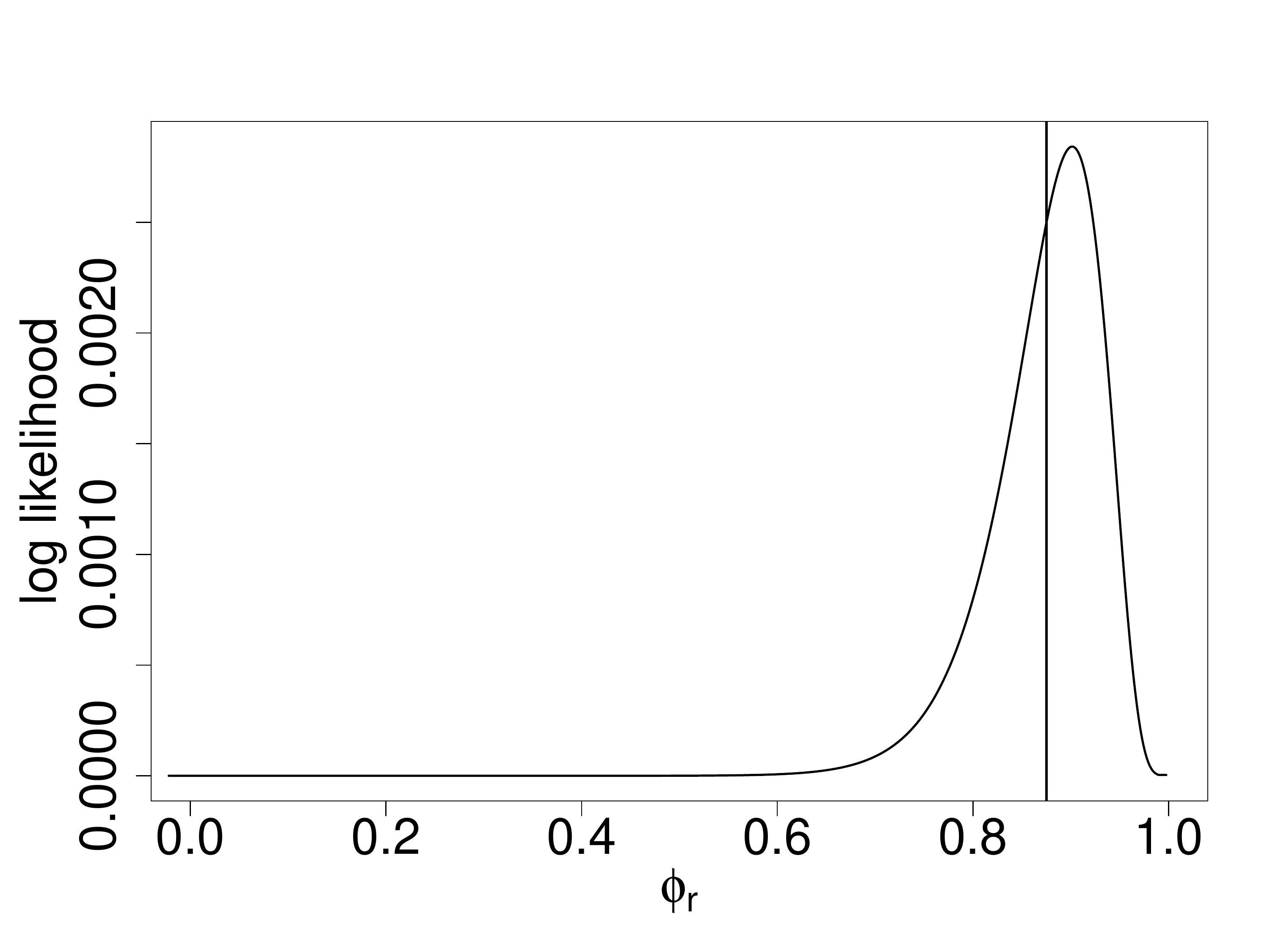}
		\caption{}
	\end{subfigure} 
	\begin{subfigure}[b]{0.328\textwidth}
		\includegraphics[width=1\textwidth]{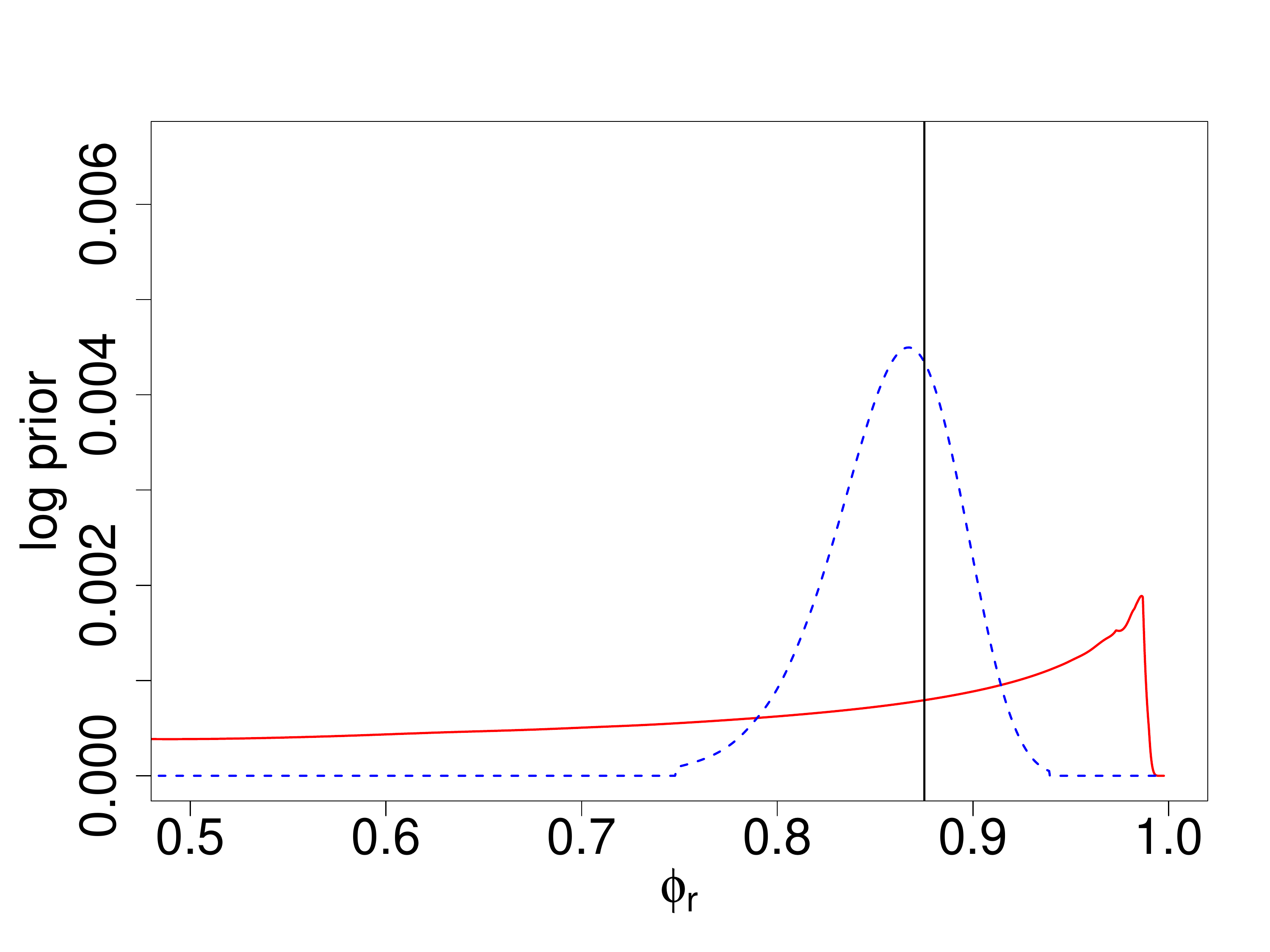}
		\caption{}
	\end{subfigure} \\
	\caption{Comparison between non-smoothing and smoothing changes in the posterior distribution of $\phi_r$ for the model \eqref{eq:ar1-ex}. (a): Scaled log posterior distributions from Step 1 (solid red) and Step 2 (dashed blue).
	(b): Scaled log likelihood function.
	(c): Scaled log prior distributions from Step 1 (solid red) and Step 2 (dashed blue). 
	The vertical line is the true value $\phi_r=0.88$.}
\label{fig_priors}
\end{figure}

\subsection{Sensitivity of prediction to smoothing} \label{sec:smooth}

There are different approaches to control the degree of smoothness in Step 2. This can be, for instance, dictated by the case study and prior knowledge. Here, we present one possible method, which is based on two metrics: the first focuses on the departure of the estimated posterior against the exact simulated distribution, and the second is based on cross-validation.

To assess the improved accuracy in capturing the true distribution of the latent process $\mb{x}_r$, for each value of $\phi_r$, we calculate the Kullback-Leibler Divergence (KL), a widely used metric for comparing two probability distributions. The departure from the true posterior $\pi(\mb{x}_r \mid \mb{y}_r, \phi_r )$ is defined as
\begin{equation}
\mbox{KL}_r = \int \tilde{\pi}(\mb{x}_r \mid \mb{y}_r, \phi_r )\log \left\{\frac{\tilde{\pi}(\mb{x}_r \mid \mb{y}_r, \phi_r )}{\pi(\mb{x}_r \mid \mb{y}_r, \phi_r )}\right\}\mathrm{d}\mb{x}_r,
\label{eq:kl}\end{equation}
where $\phi_r$ in $\tilde{\pi}$ is calculated using either Step 1 or Step 2 of our approach. A small KL$_r$ indicates a small departure from the target posterior, and a zero KL$_r$ indicates that the two distributions are the same.

The data are simulated from a known model, and the posterior distribution of the latent process $\pi(\mb{x}_r \mid \mb{y}_r, \phi_r )$ can be easily obtained from the joint distribution $\pi(\mb{x}_r, \phi_r \mid \mb{y}_r)$: 
\[
	\begin{aligned}
	\pi(\mb{x}_r \mid \mb{y}_r, \phi_r) & \propto \pi(\mb{x}_r, \phi_r \mid \mb{y}_r) \\
	& \propto \mbox{exp}  \bigg( -\frac{1}{2} \mb{x}_r^{\top} \mb{Q}_{x,r}  \mb{x}_r \bigg) \times
	\mbox{exp}  \bigg\{- \frac{1}{2} \tau (\mb{x}_r^{\top}\mb{x}_r - 2 \mb{y}_r^{\top}\mb{x}_r) \bigg\} \\
	& = \mbox{exp}  \bigg\{- \frac{1}{2} \mb{x}_r^{\top} (\mb{Q}_{x,r} + \tau \mbox{I}) \mb{x}_r + \tau\mb{y}_r^{\top}\mb{x}_r \bigg\}\\ & =\mbox{exp}  \bigg\{- \frac{1}{2} \mb{x}^{\top} \mb{P}_r \mb{x}_r + \mb{b}_r^{\top} \mb{x}_r \bigg\},
	\end{aligned}
\]
where, $\mb{P}_r = \mb{Q}_{x,r} + \tau \mbox{I}$ and $\mb{b} = \mb{y}_r^{\top} \tau$. This implies that $\pi(\mb{x}_r \mid \mb{y}_r, \phi_r )\sim \mathcal{N}_{T}(\bs{\mu}_{0,r}, \bs{\Sigma}_{0,r})$, with $\bs{\mu}_{0,r}=\mb{P}_r^{-1}\mb{b}_r$ and $\bs{\Sigma}_{0,r}=\mb{P}_r^{-1}$. We also assume that the approximated posterior in \eqref{eq:kl} is normal, i.e., $\pi_{\mbox{\scriptsize{appr}}}(\mb{x}_r \mid \mb{y}_r, \phi_r )\sim \mathcal{N}_{T}(\bs{\mu}_{1,r}, \bs{\Sigma}_{1,r})$, and we obtain $\bs{\mu}_{1,r}$ and $\bs{\Sigma}_{1,r}$ based on the sample mean vector and covariance matrix from 10,000 posterior samples. The KL divergence expression in \eqref{eq:kl} can be simplified in the case of two multivariate Gaussian distributions. Indeed, if the target distribution is $\mathcal{N}_{T}(\bs{\mu}_{0,r}, \bs{\Sigma}_{0,r})$ and the approximation is  $\mathcal{N}_{T}(\bs{\mu}_{1,r}, \bs{\Sigma}_{1,r})$, we have
\[
\mbox{KL}_r = \frac{1}{2} \bigg\{\mbox{log} \frac{|\bs{\Sigma}_{1,r}|}{|\boldsymbol{\Sigma}_{0,r}|} - T + \mbox{tr}(\boldsymbol{\Sigma}_{1,r}^{-1}\boldsymbol{\Sigma}_{0,r}) + 
(\boldsymbol{\mu}_{1,r} - \boldsymbol{\mu}_{0,r})^{\top} \boldsymbol{\Sigma}_{1,r}^{-1}(\boldsymbol{\mu}_{1,r} - \boldsymbol{\mu}_{0,r})  \bigg\},
\]
where $|\bs{\Sigma}|$ denotes the determinant of $\bs{\Sigma}$. Since the KL changes across different orders of magnitudes, we opted for a variance stabilizing estimator, the Expected Mean Log KL (EMLKL) divergence, defined as $\mbox{EMLKL}=\exp \left\{\frac{1}{R}\sum_{r=1}^R \log(\mbox{KL}_r)\right\}$

Therefore, we assess the impact of smoothing on the prediction skills of the estimated process. We use the conditional predictive ordinate (CPO) for leave-one-out cross-validation, defined as
\[
    \mbox{CPO}_r(t) = \pi\{y_r(t) | \mb{y}_r(-t)\} = \int \pi\{y_r(t) |  \mb{y}_r(-t), \theta_r\} \pi\{\theta_r | \mb{y}_r(-t)\} \mathrm{d}\theta_r,
\]
where $\mb{y}_r(-t)$ represents the vector of observations $\mb{y}_r$ with the $t$-th component omitted. In other words, $\mbox{CPO}_r(t)$ is calculated by first obtaining the predictive distribution at $t$ given all but the $t$-th observation in the time series, and then evaluating it at the actual withheld value $y_r(t)$. The CPO can be interpreted as a continuous equivalent of the posterior probability that the observation is predicted from the model, so larger values are preferable. The CPO can be computed efficiently without re-running the model $R \times T$ times \citep{held2010posterior}. The CPOs are then aggregated in an overall score for comparing different models by averaging across time and regions. As with the KL, we propose the Expected Mean Log Conditional Predictive Ordinate (EMLCPO), defined as $\mbox{EMLCPO} =\mbox{exp}\bigg[\frac{1}{RT} \sum_{r=1}^{R} \sum_{t=1}^{T} \mbox{log}\{\mbox{CPO}_r(t)\}\bigg]
$, with models having relatively higher values of EMLCPO, showing a better fit. 

\begin{figure}[htb!]
	\centering
	\begin{subfigure}[b]{0.471\textwidth}
		\includegraphics[width=1\textwidth]{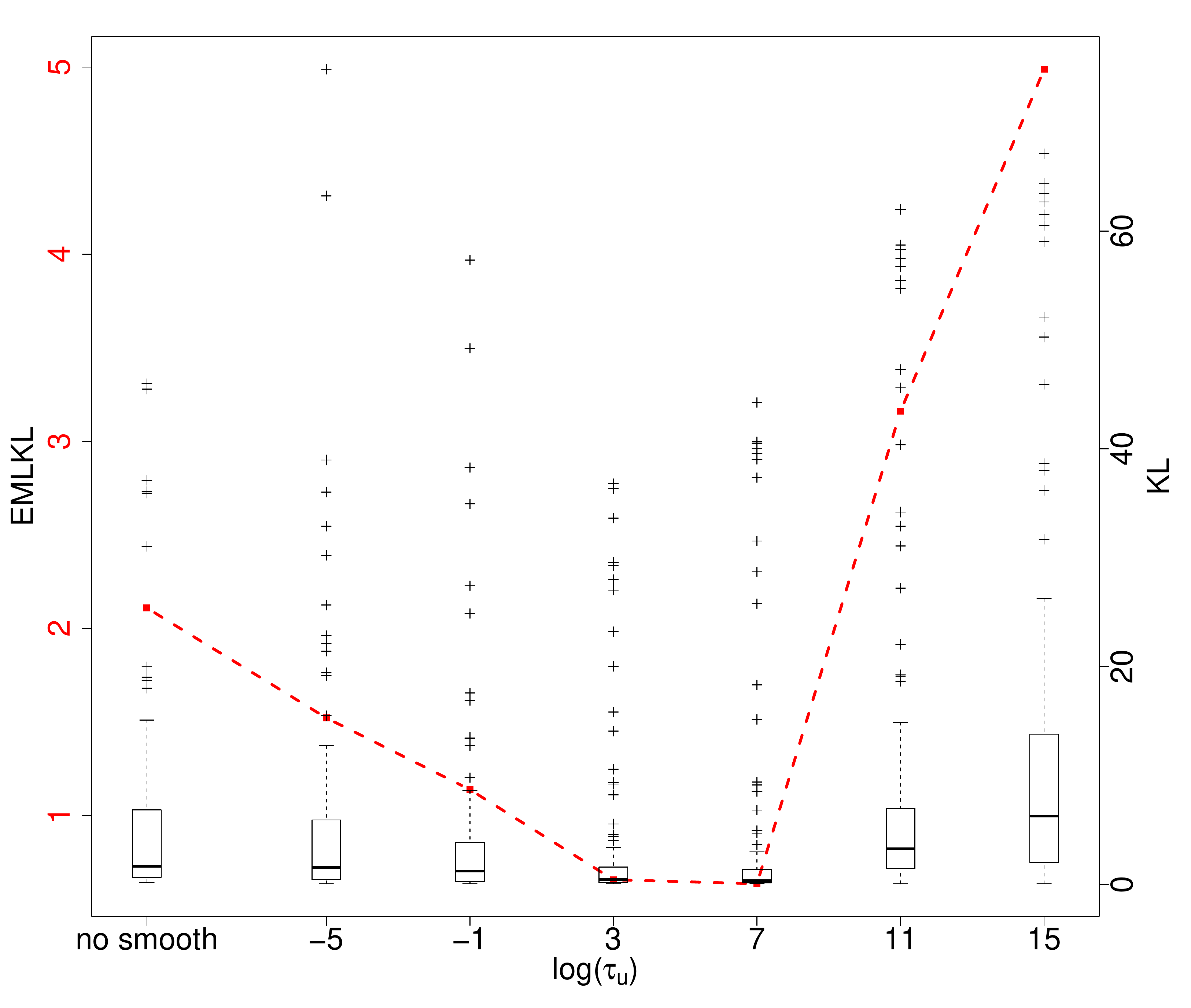}
		\caption{}
	\end{subfigure}
	\begin{subfigure}[b]{0.471\textwidth}
		\includegraphics[width=1.04\textwidth]{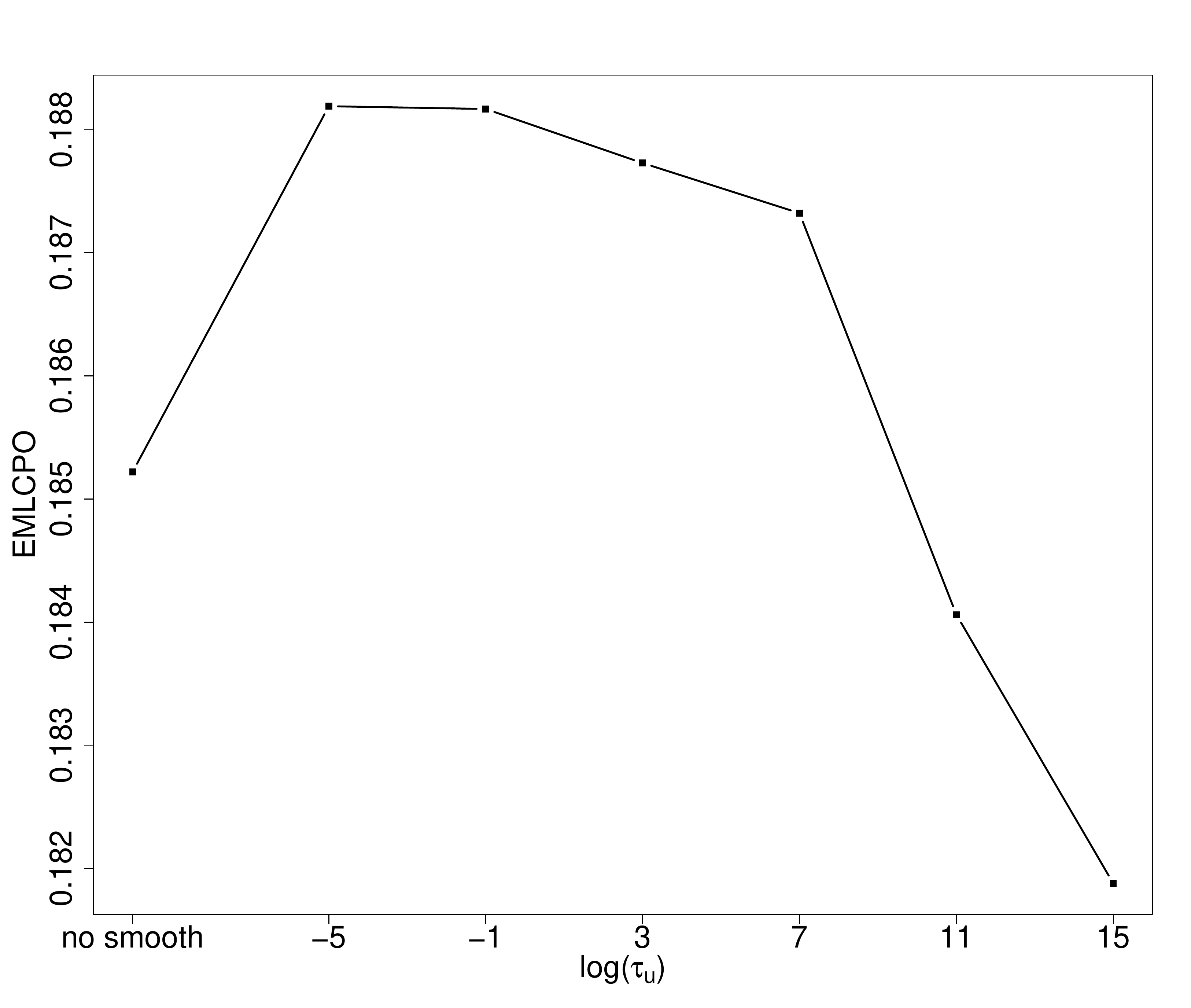}
		\caption{}
	\end{subfigure}	
	\caption{Comparison of inference and prediction performances between the standard non-smoothing method and the proposed smoothing approach from fitting \eqref{eq:ar1-ex}. (a): EMLKL (dashed red, left $y$-axis) and $\mbox{KL}_r$ (black, right $y$-axis) for no smoothing along with 6 different degrees of smoothness. Lower values of $\mbox{log}(\tau_{u})$ indicate less smoothing. (b): Corresponding EMCPO for the same degrees of smoothing.}
\label{CPO-KL-sg1}
\end{figure}

We compare the EMLKL and the EMLCPO based on six different degrees of smoothing by changing the values of $\tau_u$ in \eqref{eq:rw2}: $\mbox{log}(\tau_{u}) = \{-5, -1,  3 , 7, 11, 15\}$, with lower values of $\mbox{log}(\tau_{u})$ indicating less smoothing. 
Here, $\mbox{log}(\tau_{u})=15$ results in a constant value across the regions (complete smoothing), so no larger values are considered. Figure \ref{CPO-KL-sg1} shows the results based on (a) KL and on (b) CPO according to the various degrees of smoothing. The first value in the $x$-axis, `no smooth', corresponds to the estimates directly from Step 1 of our approach. According to the EMLKL (panel (a), left $y$-axis) and the EMLCPO (panel (b)),  the best fit occurs when $\mbox{log}(\tau_{u})=7$ and $\mbox{log}(\tau_{ u})=-5$, respectively. For the EMLKL there is a minimal difference between the $\mbox{log}(\tau_{u})=3$ and $\mbox{log}(\tau_{u})=7$, and the right $y$-axis highlights how the first choice results in less variable KL divergences. Both scores show that there is a clear improvement against a model with no smoothing for $\mbox{log}(\tau_{u})=\{-5, -1,  3 , 7\}$.
After $\mbox{log}(\tau_{u})=7$ the posteriors are oversmoothed and this worsens the fit compared to no smoothing (high EMLKL and low EMLCPO values). Evidence from this numerical study suggests that smoothing almost always improves the estimation of the latent process and prediction. The overall agreement between EMLKL and EMLCPO is essential, as, in a real application, the actual underlying distribution is unknown, and a cross-validation metric, such as the EMLCPO, would be used for choosing the optimal degree of smoothness. 

Smoothing does not just improve the prediction and decrease the bias, but also results in less variable estimates. Figure \ref{CPO-KL-sg1}a (right $y$-axis) shows the spread of KL$_r$ for the different amounts of smoothing, displayed as a boxplot. It is readily apparent that optimal smoothing results in more stable estimates by decreasing the variance across regions. 

\section{Application to the WRF data set} \label{sec:wrf}

In this section, we apply the approaches in Section~\ref{sec:method} to model and predict a sizeable simulated wind speed data set in Saudi Arabia. The predictive ability at the sub-grid scale is of interest for statistical downscaling.  Interpolated wind from the numerical simulation can be used as a baseline to build a mathematical relationship (e.g., pattern scaling) from in-situ ground wind data at the same location, hence allowing to generate more accurate, observation-driven wind maps

We apply our method to a spatial data set of simulated wind speed detailed in Section~\ref{sec:wrf_data}.
In Section~\ref{sec:model} we present the local model that is fitted to each region and in Section~\ref{sec:windresults} we explain in details each step of our approach and present the results.

\subsection{The WRF data set} \label{sec:wrf_data}

We focus on a simulation generated by \cite{yip18} from the Weather Research and Forecasting (WRF) model, which is a state-of-the-art Numerical Weather Prediction model, developed at the National Center for Atmospheric Research and also recently used in wind energy assessment in \cite{tag20}. Mesoscale numerical models such as WRF rely on large-scale atmospheric phenomena or meteorological reanalysis to provide boundary conditions and solve physical equations driving the real processes on a fine scale. The boundary conditions used to simulate the WRF data are obtained from the Modern-Era Retrospective analysis for Research and Applications (MERRA, \citet{rienecker2011merra}), a reanalysis product developed at NASA's Global Modeling and Assimilation Office, using the Goddard Earth Observing System Version 5 general circulation model, together with satellite and surface observations through a data assimilation system.
 
Each simulation corresponds to hourly data of the zonal and meridional ($U$ and $V$) wind components on a regular grid of $769\times659$ points in space (5-km resolution) bounded by 5-$35^{\circ}$N and 30-$65^{\circ}$E during the 2009-2014 period, at 2 meters above ground level. The full data set comprises of 506,771 spatial locations. We select data that fall inside Saudi Arabia from 03/06/2010 between 14:00 and 15:00 local time, when wind speeds tend to peak, resulting in 84,494 points in space. The $U$ and $V$ components are converted into wind speed: $\sqrt{U^2 + V^2}$. Figure \ref{fig:map-data-cl}a shows the map of the wind field. 

We first partition the domain into $R$ regions small enough so that the assumption of stationarity is plausible. The disjoint subsets are obtained using the $k$-means clustering method, which minimizes the sum of squares from points to the assigned region centers \citep{hartigan1979algorithm}. It is, in principle, possible to provide a more formal assessment of stationarity and use it as a metric for clustering, for example, by fitting directional variograms to each region. However, because these estimates are corrected with a smoothing step, the clustering method is less critical. Our partition results into $R=2000$ regions, (see Figure~\ref{fig:map-data-cl}b), with the smallest region containing 26 locations ($\approx 28 \times 28$ km) and the largest 62 ($\approx 48 \times 48$ km).

\subsection{The spatial model}\label{sec:model}

The distribution of wind speed is bounded below by zero and is significantly right-skewed. Therefore, wind speed cannot be directly modeled with the Gaussian distribution. Common transformations for normalizing wind speed data include logarithmic transformation and square-root transformation \citep{taylor2009wind}.
\citet{haslett1989space} showed that square-root transformation is well suited for wind data, as the resulting transformed wind speed resembles the Gaussian distribution. Hence, for each region $r$ we model the square-root transformed wind speed $y_r$ at sampling locations $\mb{s}=(\mb{s}_1, \ldots, \mb{s}_{N_r})$ with a latent Gaussian model, a special case of the hierarchical framework proposed in \eqref{eq:hierarq}. For each region $r$, we assume
\[
y_r(\mb{s}_i) =  \mb{z}_r(\mb{s}_i)^{\top} \bs{\beta}_r + u_r(\mb{s}_i) + \epsilon_r(\mb{s}_i), \quad i=1, \ldots, N_r,
\]
where $\mb{z}_r$ is a $p$-dimensional vector of covariates, and $\bs{\beta}_r$ is the vector of the linear coefficients. Here, $\{\epsilon_r(\mb{s}_1), \ldots, \epsilon_r(\mb{s}_{N_r})\} \sim \mathcal{N}_{N_r} (0, \tau_{\epsilon, r}^{-1}I_{N_r})$ is the iid random noise that accounts for the model uncertainty. The aforementioned model can be written in the vector form
\begin{equation}
\mb{y}_r | \bs{\beta}_r, \mb{u}_r, \tau_{\epsilon, r} \sim \mathcal{N}_{N_r}(\mb{Z}_r \bs{\beta}_r + \mb{u}_r, \tau_{\epsilon, r}^{-1} \mb{I}_{N_r}),
\label{eq:like-Y-2}
\end{equation}
where $\mb{y}_r = \{y(\mb{s}_1), \ldots, y(\mb{s}_{N_r})\}^{\top}$ is the observation vector and the $N_r \times P$ design matrix is $\mb{Z}_r=\{\mb{z}_r(\mb{s}_1), \ldots,\mb{z}_r(\mb{s}_{N_r})\}^\top$. We consider $p=2$, thus two covariates: elevation and distance to the coast. In terms of the hierarchical framework in Section \ref{sec:background}, \eqref{eq:like-Y-2} is the first equation, i.e., the data level, in \eqref{eq:hierarq}. 

The spatial field $u_r(\mb{s}_i)$ is assumed to be Gaussian and isotropic, with a covariance described by the Mat\'{e}rn function, a widely popular choice in spatial statistics. For two locations $\mb{s}_1$ and $\mb{s}_2$ at distance $h=\|\mb{s}_1-\mb{s}_2\|$, the Mat\'{e}rn covariance is defined as \citep{ste99}
\begin{equation}
\mbox{cov}\{u_r(\mb{s}_1),u_r(\mb{s}_2)\}=C_r(h) = \sigma^2_{u,r} \frac{1}{\Gamma(\nu_r)2^{\nu_r-1}}(\kappa_r h)^{\nu_r} {\cal{K}}_{\nu_r}(\kappa_r h),
\label{eq:matern}
\end{equation}
where $\sigma^2_{u,r}=1/\tau_{u,r}$ is the marginal variance and ${\cal{K}}_{\nu_r}$ is the modified Bessel function of the second kind of order $\nu_r>0$. The popularity of the Mat\'ern is mainly attributable to the control of the number of mean square derivatives of the underlying process through the parameter $\nu_r$. The range is controlled by $\kappa_r >0$ and $\rho_r = \sqrt{8\nu_r}/\kappa_r$ represents the distance at which the spatial correlation is approximately $0.13$, and we set $\nu_r=1$. 

The vector of hyperparameters to be estimated is given by the precision of the data, the precision of the latent process, and its range, so that 
\[
\bs{\theta}_r =(\theta_{1,r},\theta_{2,r},\theta_{3,r})^\top= \{\log( \tau_{\epsilon, r}), \log(\tau_{u,r}), \log(\rho_r)\}^\top.
\]
The linear coefficients $\bs{\beta}_r$ in \eqref{eq:like-Y-2} are less variable, so they are not included in the vector of hyperparameters to be smoothed.

We provide a joint distribution for the range $\rho_r$ and the variance $\sigma^2_{u,r}$ using the concept of the Penalized Complexity (PC) prior that was recently introduced by \citet{simpson2017penalising}. PC develops priors that allow shrinkage towards a base model, which is assumed to be the reference. The prior is then built by allowing a control of the KL divergence from the base to the actual model. Following \citet{fuglstad2019constructing}, we assume a base model with infinite range and precision, i.e., a constant, and we assign PC priors to $\rho_r$ and  $\tau_{u,r}$ that are able to control the tail probabilities: $\mbox{P}(\sigma^2_{u,r} > \sigma^2_{0,r}) = \alpha_1$ and $\mbox{P}(\rho_r < \rho_{0,r}) = \alpha_2$. We choose $\alpha_1 =\alpha_2= 0.01$, $\rho_{0,r}$ to be the $20\%$ of the range of the observations and $\sigma^2_{0,r}$ the variance estimated from the data at region $r$. In other words, we assume a prior that bounds the variance to be larger than that estimated from the data with a $1\%$ chance, and the range to be below $20\%$ of the range of the observations with a $1\%$ chance. 
For $r=1, \ldots, R$, we assume a vague Gamma prior with parameters 1 and 0.00005 for $\tau_{\epsilon, r}$ and a vague Gaussian prior $\mathcal{N}(0,1000)$ for $\bs{\beta}_r$. The priors are also assumed to be independent across components. 
The \textit{R-INLA} package is used for model fitting and predictions  \citep{rue2009approximate}.

\subsection{Results} \label{sec:windresults}

We now detail our approach with the data and the model described in the previous sections. The three steps are described as follows:

{\bf{Step 1: The model fitted to each region}}. 

\begin{figure}[htb!]
	\centering
		\begin{subfigure}[b]{0.328\textwidth}
		\includegraphics[width=\textwidth,height=5.5cm]{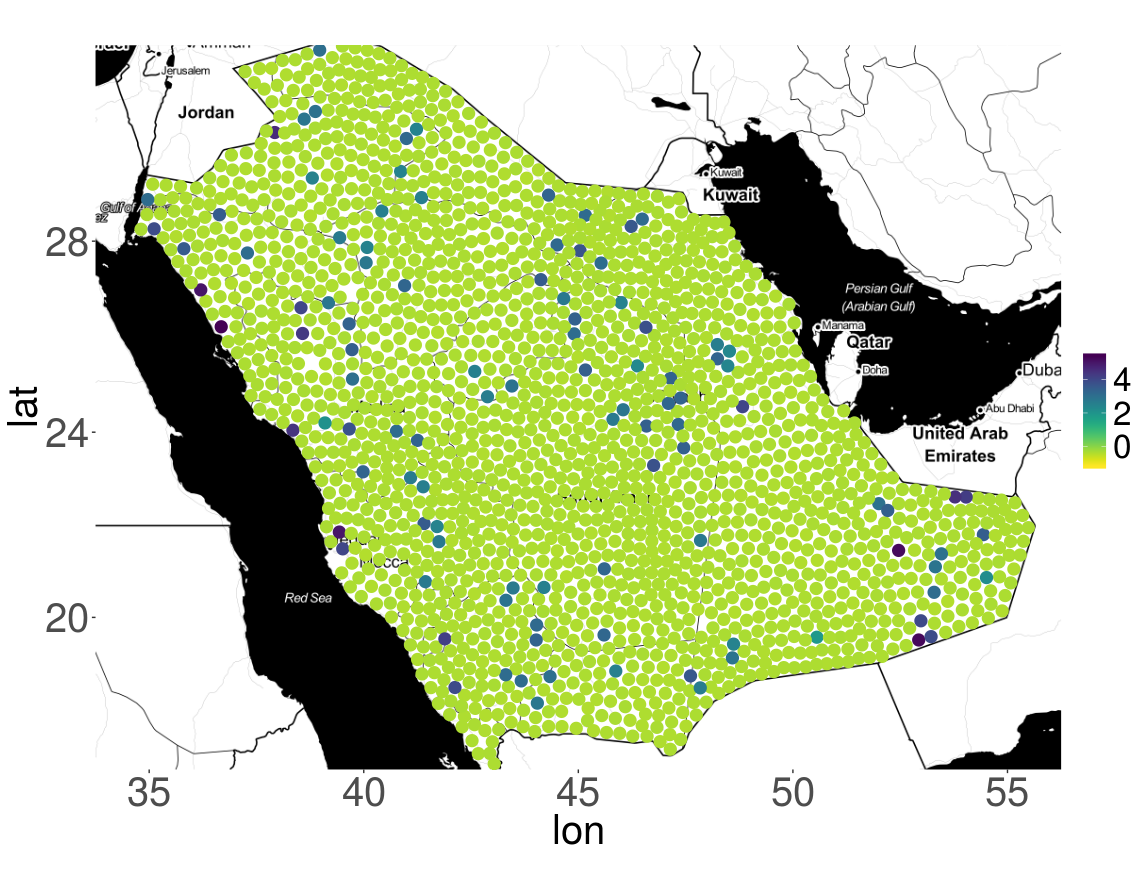}
		\caption{$\hat{\theta}_{3,r}$}
	\end{subfigure} 
	\begin{subfigure}[b]{0.328\textwidth}
		\includegraphics[width=\textwidth,height=5.5cm]{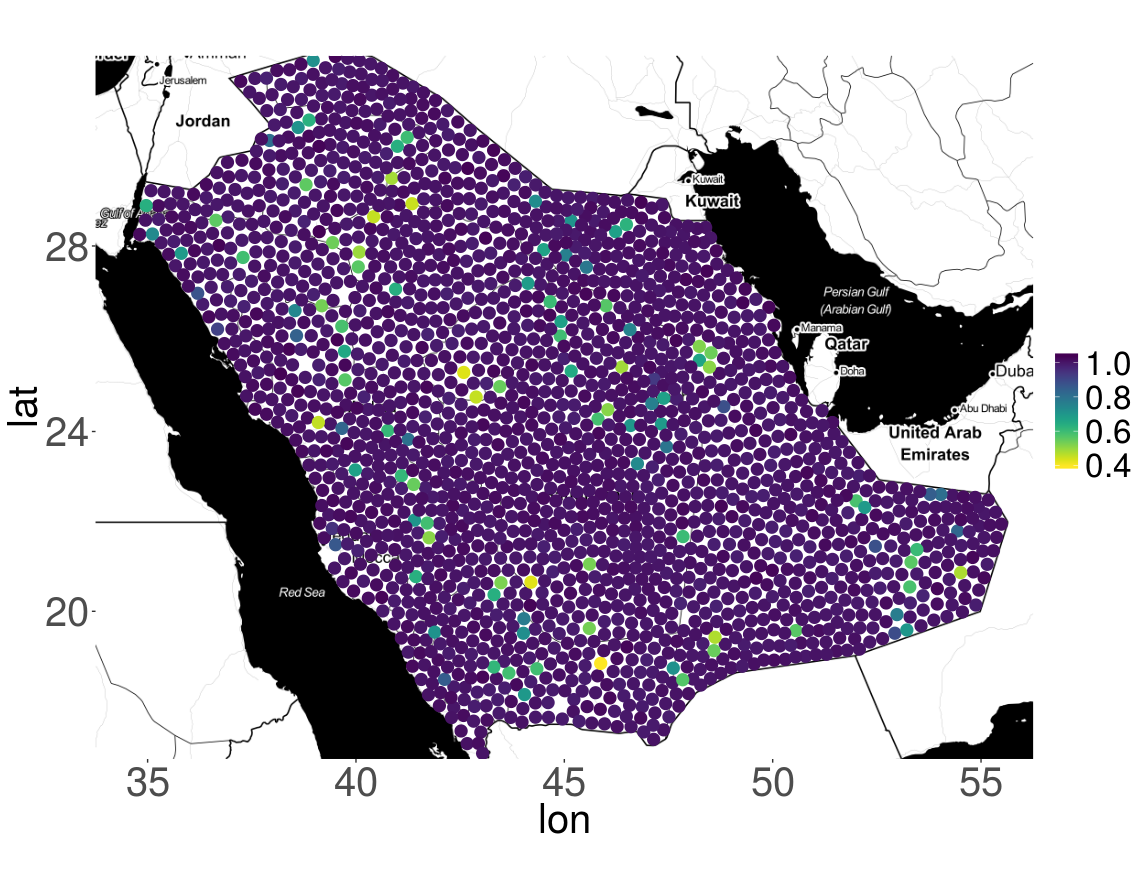}
		\caption{$\widehat{\mbox{sd}}_{3,r}$}
	\end{subfigure} 
	\begin{subfigure}[b]{0.328\textwidth}
		\includegraphics[width=\textwidth,height=5.5cm]{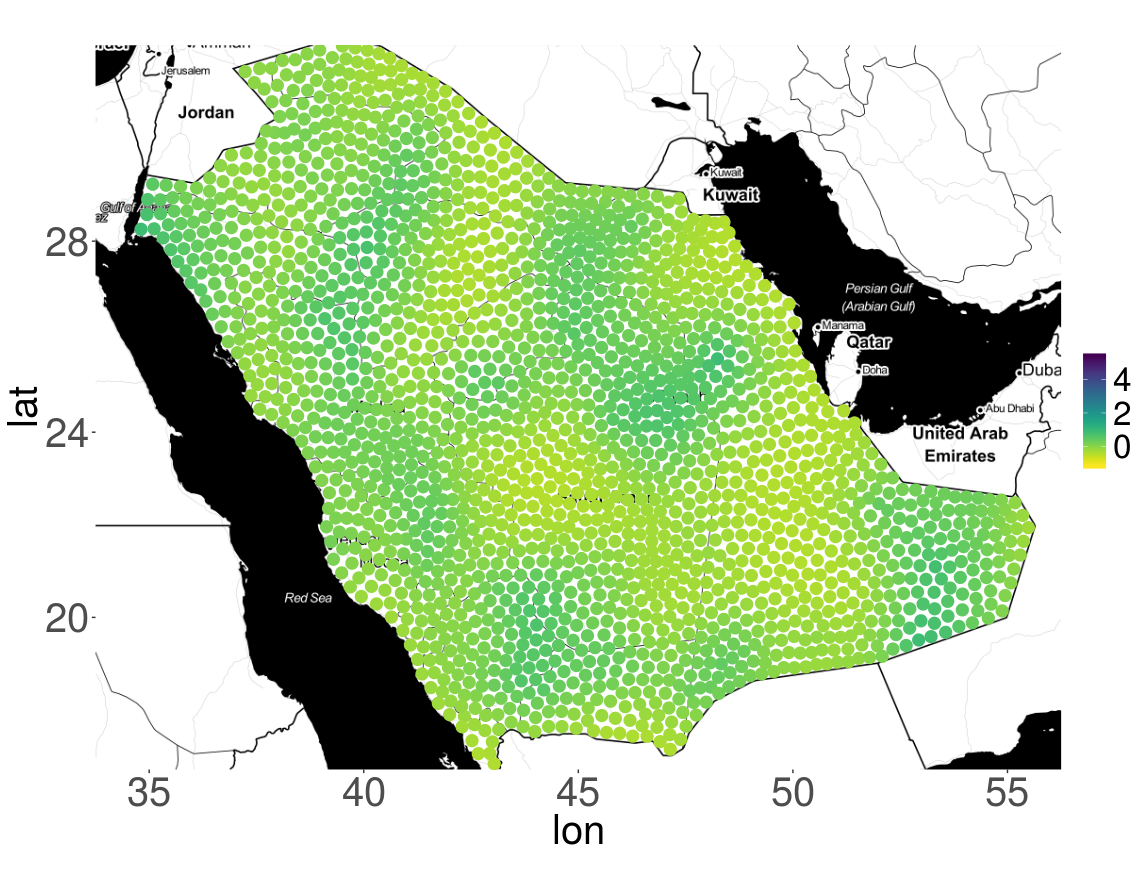}
		\caption{}
	\end{subfigure} 
		\caption{Map of the estimated (a) mode and (b) standard deviation of $\tilde{\pi}(\theta_{3,r} \mid \mb{y}_r)$ for the $R=2000$ regions shown in Figure~\ref{fig:map-data-cl}b. (c) posterior mode of $\tilde{\pi}_{\mbox{\scriptsize{smooth}}}(\theta_{3,r} \mid \mb{y})$.}
		\label{fig:map-smooth}
\end{figure}

We fit the model outlined in Section~\ref{sec:model} to each of the $R=2000$ regions in Figure~\ref{fig:map-data-cl}b separately, and obtain estimates of the posterior distribution for the $k$-th element of $\bs{\theta}_r$ for $k=1,2,3$, which we denote by $\tilde{\pi}(\theta_{k,r} \mid \mb{y}_r)$. We denote as $\hat{\theta}_{k,r}$ the mode of $\tilde{\pi}(\theta_{k,r} \mid \mb{y}_r)$, while the posterior standard deviation is denoted as $\widehat{\mbox{sd}}_{k,r}$. We show the results for $\theta_{3,r}=\log(\rho_r)$, since the range is the hardest parameter to identify, and hence the most variable across regions. Figure~\ref{fig:map-smooth}a shows the maps of $\hat{\theta}_{3,r}$. Many regions have a considerably higher estimated posterior mode than the neighboring regions, hence smoothing is necessary. Figure~\ref{fig:map-smooth}b  shows  the map of the posterior standard deviation $\widehat{\mbox{sd}}_{3,r}$, and it is apparent how the locations with large range values correspond to the ones with low posterior variance. The high variance in the estimates of Figure~\ref{fig:map-smooth} a is a consequence of the small region size needed to accommodate the non-stationarity. 
The region size is another tuning parameter of the method, and cross-validation could have been used to choose the optimal region size.

{\bf{Step 2: Smoothing the hyperparameters}}.

The modes $\hat{\theta}_{k,r}$ from Step 1 are smoothed here independently across $k$ for simplicity and are normalized by subtracting the mean and dividing by its standard deviation computed across $r=1, \ldots ,R$, for each $k$.  For a general Mat\'ern in region $r$ with known smoothness $\nu_r$, under infill asymptotics, only $\sigma_r^2 \kappa_r^{2\nu_r}$ can be consistently estimated \citep{zha04}. As the likelihood and the posteriors can be concentrated around non-linear manifolds of the parameter space, modelling the hyperparameters in the log-scale alleviates the problem of smoothing them separately.
With an abuse of notation, we now refer to $\bs{\theta}_r$ and their components as their normalized version. We assume an additive model for smoothing: $\hat{\theta}_{k,r}(\mb{s}_c) = u_{k}(\mb{s}_c) + \varepsilon_{k,r}(\mb{s}_c)$, where the locations $\mb{s}_c$ are the centroids of each region $r$. The process $u_k(\mb{s}_c)$ is assumed to be Gaussian and modeled with the Mat\'{e}rn covariance in \eqref{eq:matern}, with marginal variance $\tilde{\sigma}^2_{u,k}$, range $\tilde{\rho}_{k}$, and the iid noise is $\varepsilon_{k,r} \sim \mathcal{N} (0, \tilde{\tau}^{-1}_{\epsilon; k, r})$, for $k=1, 2$ and 3. 

We assume $\tilde{\tau}_{\epsilon; k, r}$ to be fixed at the value of $1/\widehat{\mbox{sd}}_{k,r}^2$, $r=1,\ldots, R$, from Step 1. This ensures that the same degree of smoothness is applied to all three additive models, i.e., the hyperparameters with a larger standard deviation will be smoothed more than the ones with a smaller standard deviation.
Here, $\tilde{\rho}_{k}$ is fixed to half of the domain of the study region. A choice of considerably different values, such as the size of the domain, would result in oversmoothing. The choice of $\tilde{\tau}_{u} = 1/\tilde{\sigma}^2_{u,k}$ is performed via cross-validation and will be discussed later. Because $\hat{\theta}_{k,r}, k=1,2,3$, are at the same scale after normalization, we can use the same smoothness and therefore $\tilde{\tau}_{u}$ will not be strongly dependent on $k$. We use six equally spaced values for log($\tilde{\tau}_{u}$), varying from $-7.5$ to $5$. The fitted values from the smoothing are then transformed back from the normalized to the original scale. Figure~\ref{fig:map-smooth}c  is an example of the estimated posterior mode of $\tilde{\pi}_{\mbox{\scriptsize{smooth}}}(\theta_{3,r} \mid \mb{y}_r)$ with $\mbox{log}(\tilde{\tau}_{u})=-5$.

{\bf{Step 3: Re-fit the model to each region using integration points}}. 

In the AR(1) simulated example in Section~\ref{sec:ar1}, the smoothed hyperparameter posterior was assumed to be a point mass concentrated at the smoothed posterior mode from Step 2, so that calculation of $\tilde{\pi}_{\mbox{\scriptsize{smooth}}}\{x_r(i) \mid \mb{y}_r\}$ in \eqref{eq:approxx2} was trivial. In this application, we propose a more articulated method which numerically approximates the integral in the first equation of \eqref{eq:approxx2}. 

We use the Gauss–Hermite quadrature, a numerical scheme to approximate integrals of the form $\int e^{-\xi^2} f(\xi) \mathrm{d}\xi \approx \sum_{l=1}^{L} f\{\xi^{(l)}\} \Delta^{(l)}$ for a fixed $L$. The abscissas for the quadrature of order $L$, which are given by the roots of the Hermite polynomials $\xi^{(l)}$, and the weights $\Delta^{(l)}$, both have a closed form expression \citep{abr64}.

We operate under the assumption that $\tilde{\pi}_{\mbox{\scriptsize{smooth}}}(\bs{\theta}_{r} \mid \mb{y})$ can be well approximated by a product of marginal normal distributions $\tilde{\pi}_{\mbox{\scriptsize{smooth}}}(\bs{\theta}_{r} \mid \mb{y})\approx \prod_{k=1}^3\mathcal{N}(\mu_{k,r},\sigma^2_{k,r})$, where $\mu_{k,r}$ and $\sigma^2_{k,r}$ are the posterior mean and variance of $\tilde{\pi}_{\text{smooth}}({\theta}_{k,r} \mid \mb{y})$, respectively. The independence implied by the product is made for convenience, although empirically, we found a relatively low correlation between the components of $\bs{\theta}_{r}$.
Also, because of the log-scale, the posteriors can be well approximated by a Gaussian distribution, and the first expression in \eqref{eq:approxx2} becomes 
\begin{eqnarray}
\tilde{\pi}_{\mbox{\scriptsize{smooth}}}\{x_r(i) \mid \mb{y}_r\}  & = & \int \tilde{\pi}_{\mbox{\scriptsize{smooth}}}(\mb{x}_r \mid \mb{y}_r, \bs{\theta}_{r}) \prod_{k=1}^3 \frac{1}{\sqrt{2\pi}\sigma^2_{k,r}}\exp\left\{-\frac{(\theta_{k,r}-\mu_{k,r})^2}{2\sigma^2_{k,r}} \right\} \mbox{d} \bs{\theta}_{r}  \nonumber\\
& = & \frac{1}{\sqrt{\pi}}\int \tilde{\pi}_{\mbox{\scriptsize{smooth}}}\left(\mb{x}_r \mid \mb{y}_r, 
\sum_{k=1}^3 \mu_{k,r} + \sqrt{2}\xi_{k,r}\sigma_{k,r}\right) \exp\left( -\sum_{k=1}^3 \xi_{k,r}^2 \right) \mbox{d} \xi_{1,r}  \mbox{d} \xi_{2,r}  \mbox{d} \xi_{3,r}  \nonumber\\
& \approx & \frac{1}{\sqrt{\pi}}  \sum_{l_1=1}^L \sum_{l_2=1}^L \sum_{l_3=1}^L  \tilde{\pi}_{\mbox{\scriptsize{smooth}}}\left(\mb{x}_r \mid \mb{y}_r, \sum_{k=1}^3 \mu_{k,r} + \sqrt{2}\xi^{(l_k)}_{r}\sigma_{k,r} \right) \Delta^{(l_1)} \Delta^{(l_2)} \Delta^{(l_3)},\nonumber
\end{eqnarray}
where the latent field $\mb{x}_r = (\mb{u}^\top_r, \bs{\beta}^\top_r )^\top$ contains the linear coefficients and the spatial process in (\ref{eq:like-Y-2}).
Using a change of variables, we obtain $\xi_{k,r}=\frac{\theta_{k,r}-\mu_{k,r}}{\sqrt{2}\sigma_{k,r}} \Leftrightarrow \theta_{k,r} = \mu_{k,r} + \sqrt{2}\xi_{k,r}\sigma_{k,r}$. 
For this case study, $L=5$ integration points in each of the three dimensions provide an approximation that is sufficiently accurate. Thus, the required number of configurations to evaluate the integral is $L^3=5^3 = 125$. Since each configuration can be evaluated independently, the computations can be easily parallelized.

\subsection{Choice of the smoothing parameter}

\begin{figure}[htb!]
	\centering
		\includegraphics[width=13cm]{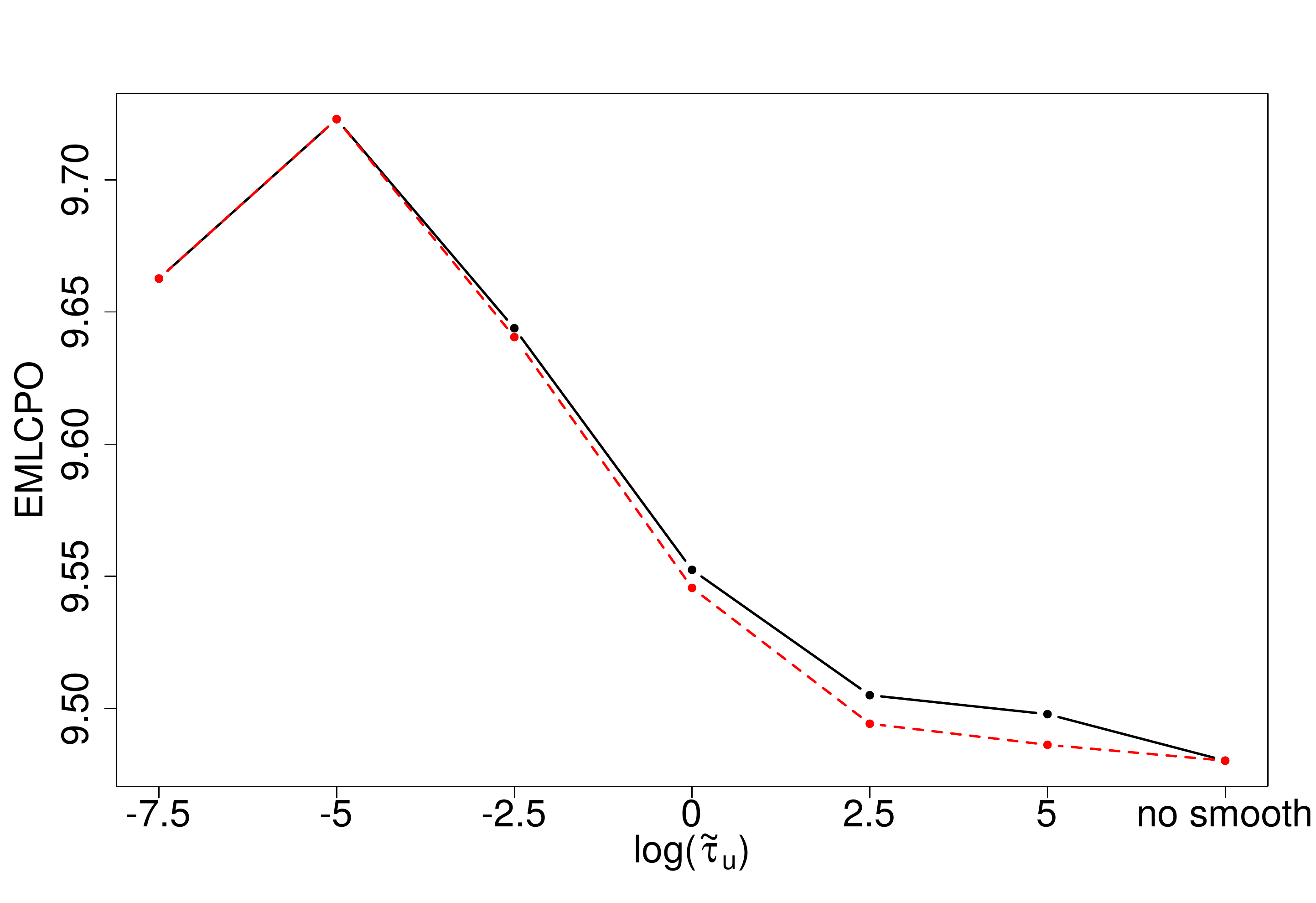}
		\caption{EMLCPO values for 6 different degrees of smoothness as well as no smoothing, the last being the results directly from Step 1. The dashed red indicates marginal posteriors computed with a point mass and the black solid with the Gauss-Hermite quadrature.
		From left to right: very smooth to no smoothing.}
		\label{fig:wrf_cpos_6smo}
\end{figure}

There is no true underlying model here, so the EMLKL in Section \ref{sec:smooth} is not applicable and we only focus on the cross-validation score EMLCPO. We compare the leave-one-out predictive performance using the different degrees of smoothing, as explained previously in Section 3. Figure~\ref{fig:wrf_cpos_6smo} shows this comparison: lower values of $\mbox{log}(\tilde{\tau}_u)$ indicate more smoothing than higher values. The highest value corresponds to the results obtained directly from Step 1. The EMLCPO value attains its maximum at $\mbox{log}(\tilde{\tau}_u) = -5$, and any of the smoothing levels improves the original estimates from Step 1. Differently from the AR(1) case in Section~\ref{sec:ar1}, where at some point the smoothing becomes excessive and the scores progressively deteriorate, here the performance is significantly improved even for a large smoothing. We also compare the predictive performances of the integration method against the approach using only the mode as in the AR(1) case. The Gauss-Hermite integration shows marginal improvement, especially for low degrees of smoothing. 
For higher degrees of smoothing, the estimated posterior distribution is more narrow, and the effect of the integration is less apparent.

\section{Discussion} \label{sec:disc}

In this work, we developed a new three-step approach for analyzing large data sets with spatial dependence that improves local models in terms of inference and prediction. The method is scalable to extremely large spatial data sets and can properly propagate the uncertainty across steps in a Bayesian framework. In Step 1, the domain is partitioned into regions, and local models are fit to each region. The size of these regions is a bias-variance trade-off; larger regions will have a smaller variance and more substantial bias, whereas smaller regions will have higher variance and lower bias. We choose to use smaller regions, thus allowing the capture of local non-stationarities, followed by a correction for the high variance, based on borrowing information from neighboring regions in Step 2, while accounting for the uncertainty of the parameter estimates from Step 1. Finally, in Step 3, the model is re-fitted to each region, propagating the uncertainty from the smoothing back into the analysis as the new posterior, thus avoiding problems of using the data twice. The approach allows flexible modeling of complex dependence structures, but is at the same time computationally affordable, as the proposed adjustment is amenable to full parallelization across regions.

In both the AR(1) simulated data and the application, the improvement from our method compared to fitting local models to each region is apparent. Indeed, the smoothing adjustment allows us to better recover the actual posterior distribution in the simulation study, and most importantly, it enables a superior predictive skill. The smoothing can be chosen to achieve the best possible advantage over the uncorrected model. Ad-hoc sensitivity analysis shows that our method is robust concerning the smoothing technique, with improved results for a wide range of smoothings. The existing methods for the nonstationary case involve highly complex model fitting strategies, and only a few software is available, see, e.g., \citet{gramacy2007tgp} and \citet{risser2015local}. However, a detailed comparison with other methods was not feasible due to the large size of our simulated wind data set, which could not be presently handled by either of these packages.

Our method is general and can be applied to many settings: space, time, space/time, and different domains, as long as a partition is provided. It relies on local models defined through a hierarchical latent process framework, a class large enough to allow a wide range of applications. 
If better local models are provided, our method can still be used to correct the variance of the estimated parameters.

A limitation of this approach lies in the assumption of a domain partition. For some applications such as wind, the regions imply a discontinuity at the border, and hence prediction at unsampled locations at the border might be suboptimal. Partition-based approaches that do not imply independence across blocks of the partition with a globally valid model are available. \cite{gui13} and \cite{cas17} proposed an evolutionary spectrum model to capture high-frequency temperature across day and night regimes. If the nature of the problem suggests a change in spatial dependence dictated by some geographical features such as mountain range as proposed in \cite{jeo18}, then this strategy could be naturally employed with appropriate likelihood approximation. For our domain and wind, however, the partition must be provided by a clustering scheme such as the $k$-means. 

An application of our model to spatio-temporal data is possible. Still, it would likely require additional approximations and a careful choice of the regions as the data size and the hyperparameter space will be considerably larger. 

\clearpage
\newpage

\appendix

\section*{Appendix: Retrieving the priors}

The re-fitting procedure in Step 3 of our approach uses the information from Step 2 as the new posterior distribution. 
We show how to retrieve the prior distribution that corresponds to the posterior for the toy example in Section~\ref{sec:ar1}.

For each $\phi_r$ and corresponding data $\mb{y}_r$, with $r=1, \ldots, R$, let $ \pi(\mb{y}_r \mid \phi_r)$ be the likelihood of observing data $\mb{y}_r$ given the hyperparameter $\phi_r$. 
We denote by $\tilde{\pi}(\phi_r \mid \mb{y}_r)$ and $\tilde{\pi}_{\mbox{\scriptsize{smooth}}}(\phi_r \mid \mb{y}_r)$ the posterior distributions from Steps 1 and 3, respectively, and $\tilde{\pi}(\phi_r)$ and $\tilde{\pi}_{\mbox{\scriptsize{smooth}}}(\phi_r)$ are the corresponding priors.

From Bayes' Theorem, the prior distributions are given by: 
\begin{equation}
    \begin{aligned}
    \label{priors}
\mbox{log}\{\tilde{\pi}(\phi_r)\} & = A + \mbox{log}\{\tilde{\pi}(\phi_r \mid \mb{y}_r )\} - \mbox{log}\{\pi(\mb{y}_r \mid \phi_r),\} \\
\mbox{log}\{\tilde{\pi}_{\mbox{\scriptsize{smooth}}}(\phi_r)\} & = A + \mbox{log}\{\tilde{\pi}_{\mbox{\scriptsize{smooth}}}(\phi_r \mid \mb{y}_r )\} - \mbox{log}\{\pi(\mb{y}_r \mid \phi_r)\},
    \end{aligned}
\end{equation}
where $A$ is the normalizing constant.

Recall that the posteriors $\tilde{\pi}(\phi_r \mid \mb{y}_r )$ and $\tilde{\pi}_{\mbox{\scriptsize{smooth}}}(\phi_r \mid \mb{y}_r )$ in the right hand sides of (\ref{priors}), are readily available from Steps 1 and 2, respectively.
Therefore, to evaluate $\tilde{\pi}(\phi_r)$ and $\tilde{\pi}_{\mbox{\scriptsize{smooth}}}(\phi_r)$, what remains to be computed is the likelihood term $\pi(\mb{y}_r \mid \phi_r)$, which is the same in both equations given in (\ref{priors}). To compute this term, we start by writing:
\begin{equation}
\pi(\mb{y}_r \mid \phi_r ) = \frac{\pi(\mb{y}_r, \mb{x}_r \mid \phi_r)}{\pi(\mb{x}_r \mid \mb{y}_r, \phi_r)},
\label{like0}
\end{equation}
and then compute (\ref{like0}) in two steps:
\begin{enumerate}
	\item The joint distribution $\pi(\mb{y}_r, \mb{x}_r \mid \phi_r):$
	
We assume that the marginal distribution of $x_r(1)$ is Gaussian with mean zero and variance $1/(1-\phi_r^2)$. Then, we can express the joint distribution of $\mb{x}_r$, $\pi(\mb{x}_r \mid \phi_r) = \pi\{x_r(1)\} \pi\{x_r(2) \mid x_r(1)\}, \ldots ,\pi\{x_r(T) \mid x_r(T-1)\}$, as 
\begin{equation}
\pi(\mb{x}_r \mid \phi_r) \sim \mathcal{N}_{T}(\mb{0},\mb{Q}_{x,r}),
\end{equation}
where $\mb{Q}_{x,r}$ is the tridiagonal precision matrix of an AR(1) process 
$$
\mb{Q}_{x,r} = 
\begin{pmatrix}
1&-\phi_r&&&&\\
\phi_r&1+\phi_r^2&-\phi_r&&&\\
&\cdots&\cdots&\cdots&& \\
&&&-\phi_r&1+\phi_r^2&-\phi_r\\
&&&&-\phi_r&1
\end{pmatrix}.
$$
It follows that the joint posterior distribution is 
\begin{equation}
\begin{aligned}
    \pi(\mb{x}_r, \phi_r \mid \mb{y}_r) & \propto \pi(\phi_r) \pi(\mb{x}_r \mid  \phi_r) \prod_{t =1}^{T} \pi\{y_{r}(t) \mid x_{r}(t), \phi_r\} \\
    & \propto \pi(\phi_r) |\mb{Q}_{x,r}|^{1/2} \tau^{1/2} \mbox{exp}  \bigg[ -\frac{1}{2}  \bigg\{ \mb{x}_r^{\top} \mb{Q}_{x,r}  \mb{x}_r + \tau (\mb{y}_r - \mb{x}_r)^{\top}  (\mb{y}_r - \mb{x}_r) \bigg\} \bigg].
    	\end{aligned}
    	\label{eq:jointAR1}
\end{equation}

\item The conditional distribution $\pi(\mb{x}_r \mid \mb{y}_r, \phi_r ):$

We use the fact that the conditional distribution of $\mb{x}_r$ is just the joint distribution between $\mb{x}_r$ and $\mb{y}_r$, without the terms that do not depend on $\mb{x}_r$ since $\mb{y}_r$ and $\phi_r$ are fixed:
	\begin{equation}
	\begin{aligned}
	\pi(\mb{x}_r \mid \mb{y}_r, \theta_r) & \propto \pi(\mb{y}_r, \mb{x} \mid \phi_r ) \\
	& \propto \mbox{exp}  \bigg( -\frac{1}{2} \mb{x}_r^{\top} \mb{Q}_{x,r}  \mb{x}_r \bigg) \times
	\mbox{exp}  \bigg\{- \frac{1}{2} \tau (\mb{x}_r^{\top}\mb{x}_r - 2 \mb{y}_r^{\top}\mb{x}_r) \bigg\} \\
	& = \mbox{exp}  \bigg\{- \frac{1}{2} \mb{x}_r^{\top} (\mb{Q}_{x,r} + \tau \mbox{I})  \mb{x}_r + \tau \mb{y}_r^{\top}\mb{x}_r \bigg \}.
	\end{aligned}
	\label{den}
	\end{equation}
Using the canonical form of the multivariate Gaussian distribution, we can write:
\[ \pi(\mb{x}_r \mid \mb{y}_r, \phi_r ) \propto \mbox{exp}  \bigg(- \frac{1}{2} \mb{x}_r^{\top} \mb{P}_r \mb{x}_r + \mb{b}_r^{\top} \mb{x}_r \bigg),  \]
where, $\mb{P}_r = \mb{Q}_{x,r} + \tau \mbox{I}$ and $\mb{b}_r = \mb{y}_r^{\top} \tau$. It follows that:
\[ \mb{x}_r \mid \mb{y}_r, \phi_r \sim \mathcal{N}_{T}(\mb{P}_r^{-1}\mb{b}_r, \mb{P}_r). \]
\end{enumerate}

Finally, from (\ref{eq:jointAR1}) and (\ref{den}), we can write $\pi(\mb{y}_r\mid \phi_r)$ in (\ref{like0}) evaluated at $\mb{x}_r = 0$ as
	\begin{equation}
	\begin{aligned}
\pi(\mb{y}_r \mid \phi_r )\bigg\rvert_{\mb{x}_r = 0} & \propto 
		 \frac{ |\mb{Q}_{x,r}|^{1/2} \mbox{exp}  \bigg(- \frac{1}{2} \tau \mb{y}_r^{\top}\mb{y}_r\bigg)}{|\mb{P}_r|^{1/2}\mbox{exp}  \bigg\{- \frac{1}{2}(-\mb{P}_r^{-1}\mb{b}_r)^{\top}\mb{P}_r(-\mb{P}_r^{-1}\mb{b}_r)\bigg\}} \\
		&  = \frac{ |\mb{Q}_{x,r}|^{1/2}} {|\mb{P}_r|^{1/2} \mbox{exp}  \bigg\{- \frac{1}{2}(   \mb{b}_r^{\top} \mb{P}_r^{-1}\mb{b}_r - \tau \mb{y}_r^{\top}\mb{y}_r) \bigg\}}.
	\end{aligned}
	\label{like}
	\end{equation}
	
Next, from the posteriors $\tilde{\pi}(\phi_r \mid \mb{y}_r )$ and $\tilde{\pi}_{\mbox{\scriptsize{smooth}}}(\phi_r \mid \mb{y}_r )$ on the right hand side of (\ref{priors}) that are computed in Steps 1 and 2, respectively, together with the likelihood term in (\ref{like}), we can obtain the corresponding priors in (\ref{priors}).
The right hand side plot of Figure~\ref{fig_priors} shows these exact scaled log prior distributions.

\bibliographystyle{apalike}
\bibliography{sample}

\begin{thebibliography}{}

\bibitem[Abramowitz and Stegun, 1964]{abr64}
Abramowitz, M. and Stegun, I.~A. (1964).
\newblock {\em Handbook of Mathematical Functions with Formulas, Graphs, and
  Mathematical Tables}.
\newblock Dover, New York.
\newblock Equation 25.4.46.

\bibitem[Bakka et~al., 2019]{bakka2019non}
Bakka, H., Vanhatalo, J., Illian, J.~B., Simpson, D., and Rue, H. (2019).
\newblock Non-stationary {G}aussian models with physical barriers.
\newblock {\em Spatial Statistics}, 29:268--288.

\bibitem[Bolin and Lindgren, 2011]{bolin2011spatial}
Bolin, D. and Lindgren, F. (2011).
\newblock Spatial models generated by nested stochastic partial differential
  equations, with an application to global ozone mapping.
\newblock {\em The Annals of Applied Statistics}, 5(1):523--550.

\bibitem[Castruccio and Guinness, 2017]{cas17}
Castruccio, S. and Guinness, J. (2017).
\newblock An evolutionary spectrum approach to incorporate large-scale
  geographical descriptors on global processes.
\newblock {\em Journal of the Royal Statistical Society: Series C (Applied
  Statistics)}, 66(2):329--344.

\bibitem[Damian et~al., 2001]{damian2001bayesian}
Damian, D., Sampson, P.~D., and Guttorp, P. (2001).
\newblock Bayesian estimation of semi-parametric non-stationary spatial
  covariance structures.
\newblock {\em Environmetrics}, 12(2):161--178.

\bibitem[Edwards et~al., 2019]{edw19}
Edwards, M., Castruccio, S., and Hammerling, D. (2019).
\newblock Marginally parametrized spatio-temporal models and stepwise maximum
  likelihood estimation.
\newblock arxiv.org/abs/1806.11388.

\bibitem[Fuglstad and Castruccio, 2020]{gei20}
Fuglstad, G.-A. and Castruccio, S. (2020).
\newblock Compression of climate simulations with a nonstationary global
  spatio-temporal {SPDE} model.
\newblock {\em Annals of Applied Statistics}.
\newblock in press.

\bibitem[Fuglstad et~al., 2015a]{fuglstad2015exploring}
Fuglstad, G.-A., Lindgren, F., Simpson, D., and Rue, H. (2015a).
\newblock Exploring a new class of non-stationary spatial {G}aussian random
  fields with varying local anisotropy.
\newblock {\em Statistica Sinica}, 25(1):115--133.

\bibitem[Fuglstad et~al., 2015b]{fuglstad2015does}
Fuglstad, G.-A., Simpson, D., Lindgren, F., and Rue, H. (2015b).
\newblock Does non-stationary spatial data always require non-stationary random
  fields?
\newblock {\em Spatial Statistics}, 14:505--531.

\bibitem[Fuglstad et~al., 2019]{fuglstad2019constructing}
Fuglstad, G.-A., Simpson, D., Lindgren, F., and Rue, H. (2019).
\newblock Constructing priors that penalize the complexity of {G}aussian random
  fields.
\newblock {\em Journal of the American Statistical Association},
  114(525):445--452.

\bibitem[Gramacy, 2007]{gramacy2007tgp}
Gramacy, R.~B. (2007).
\newblock tgp: an {R} package for {B}ayesian nonstationary, semiparametric
  nonlinear regression and design by treed gaussian process models.
\newblock {\em Journal of Statistical Software}, 19(9):6.

\bibitem[Guinness and Stein, 2013]{gui13}
Guinness, J. and Stein, M.~L. (2013).
\newblock Interpolation of nonstationary high frequency spatial-temporal
  temperature data.
\newblock {\em Annals of Applied Statistics}, 7(3):1684--1708.

\bibitem[Hammerling et~al., 2012]{hammerling2012mapping}
Hammerling, D.~M., Michalak, A.~M., and Kawa, S.~R. (2012).
\newblock Mapping of {C02} at high spatiotemporal resolution using satellite
  observations: Global distributions from {C02}.
\newblock {\em Journal of Geophysical Research: Atmospheres}, 117(D6).

\bibitem[Hartigan and Wong, 1979]{hartigan1979algorithm}
Hartigan, J.~A. and Wong, M.~A. (1979).
\newblock Algorithm {AS} 136: A k-means clustering algorithm.
\newblock {\em Journal of the Royal Statistical Society. Series C (Applied
  Statistics)}, 28(1):100--108.

\bibitem[Haslett and Raftery, 1989]{haslett1989space}
Haslett, J. and Raftery, A.~E. (1989).
\newblock Space-time modelling with long-memory dependence: Assessing
  {I}reland's wind power resource.
\newblock {\em Journal of the Royal Statistical Society. Series C (Applied
  Statistics)}, 38(1):1--50.

\bibitem[Held et~al., 2010]{held2010posterior}
Held, L., Schr{\"o}dle, B., and Rue, H. (2010).
\newblock Posterior and cross-validatory predictive checks: a comparison of
  mcmc and inla.
\newblock In {\em Statistical Modelling and Regression Structures}, pages
  91--110. Springer.

\bibitem[Higdon, 1998]{higdon1998process}
Higdon, D. (1998).
\newblock A process-convolution approach to modelling temperatures in the north
  atlantic ocean.
\newblock {\em Environmental and Ecological Statistics}, 5(2):173--190.

\bibitem[Hildeman et~al., 2019]{hildeman2019spatial}
Hildeman, A., Bolin, D., and Rychlik, I. (2019).
\newblock Spatial modeling of significant wave height using stochastic partial
  differential equations.
\newblock {\em arXiv preprint arXiv:1903.06296}.

\bibitem[Ingebrigtsen et~al., 2014]{ingebrigtsen2014spatial}
Ingebrigtsen, R., Lindgren, F., and Steinsland, I. (2014).
\newblock Spatial models with explanatory variables in the dependence
  structure.
\newblock {\em Spatial Statistics}, 8:20--38.

\bibitem[Jeong et~al., 2018]{jeo18}
Jeong, J., Castruccio, S., Crippa, P., and Genton, M.~G. (2018).
\newblock Reducing storage of global wind ensembles with stochastic generators.
\newblock {\em Annals of Applied Statistics}, 12(1):490--509.

\bibitem[Kuusela and Stein, 2018]{kuusela2018locally}
Kuusela, M. and Stein, M.~L. (2018).
\newblock Locally stationary spatio-temporal interpolation of argo profiling
  float data.
\newblock {\em Proceedings of the Royal Society A}, 474(2220):20180400.

\bibitem[Lindgren and Rue, 2008]{lindgren2008second}
Lindgren, F. and Rue, H. (2008).
\newblock On the second-order random walk model for irregular locations.
\newblock {\em Scandinavian Journal of Statistics}, 35(4):691--700.

\bibitem[Lindgren et~al., 2011]{lindgren2011explicit}
Lindgren, F., Rue, H., and Lindstr{\"o}m, J. (2011).
\newblock An explicit link between {G}aussian fields and {G}aussian {M}arkov
  random fields: the stochastic partial differential equation approach.
\newblock {\em Journal of the Royal Statistical Society: Series B (Statistical
  Methodology)}, 73(4):423--498.

\bibitem[Neto et~al., 2014]{neto2014accounting}
Neto, J. H.~V., Schmidt, A.~M., and Guttorp, P. (2014).
\newblock Accounting for spatially varying directional effects in spatial
  covariance structures.
\newblock {\em Journal of the Royal Statistical Society: Series C (Applied
  Statistics)}, 63(1):103--122.

\bibitem[Nychka et~al., 2018]{nychka2018modeling}
Nychka, D., Hammerling, D., Krock, M., and Wiens, A. (2018).
\newblock Modeling and emulation of nonstationary {G}aussian fields.
\newblock {\em Spatial Statistics}, 28:21--38.

\bibitem[Nychka et~al., 2002]{nychka2002multiresolution}
Nychka, D., Wikle, C., and Royle, J.~A. (2002).
\newblock Multiresolution models for nonstationary spatial covariance
  functions.
\newblock {\em Statistical Modelling}, 2(4):315--331.

\bibitem[Rienecker et~al., 2011]{rienecker2011merra}
Rienecker, M.~M., Suarez, M.~J., Gelaro, R., Todling, R., Bacmeister, J., Liu,
  E., Bosilovich, M.~G., Schubert, S.~D., Takacs, L., Kim, G.-K., et~al.
  (2011).
\newblock Merra: {NASA}’s modern-era retrospective analysis for research and
  applications.
\newblock {\em Journal of Climate}, 24(14):3624--3648.

\bibitem[Risser and Calder, 2017]{risser2015local}
Risser, M. and Calder, C. (2017).
\newblock Local likelihood estimation for covariance functions with
  spatially-varying parameters: The convo{S}{P}{A}{T} package for {R}.
\newblock {\em Journal of Statistical Software, Articles}, 81(14):1--32.

\bibitem[Risser, 2016]{risser2016nonstationary}
Risser, M.~D. (2016).
\newblock Nonstationary spatial modeling, with emphasis on process convolution
  and covariate-driven approaches.
\newblock {\em arXiv preprint arXiv:1610.02447}.

\bibitem[Rue and Held, 2005]{rue2005gaussian}
Rue, H. and Held, L. (2005).
\newblock {\em {G}aussian {M}arkov {R}andom {F}ields: {T}heory and
  {A}pplications}.
\newblock Chapman and Hall/CRC.

\bibitem[Rue et~al., 2009]{rue2009approximate}
Rue, H., Martino, S., and Chopin, N. (2009).
\newblock Approximate {B}ayesian inference for latent {G}aussian models by
  using integrated nested {L}aplace approximations.
\newblock {\em Journal of the Royal Statistical Society: Series B (Statistical
  Methodology)}, 71(2):319--392.

\bibitem[Rue et~al., 2017]{rue2017bayesian}
Rue, H., Riebler, A., S{\o}rbye, S.~H., Illian, J.~B., Simpson, D.~P., and
  Lindgren, F.~K. (2017).
\newblock Bayesian computing with {I}{N}{L}{A}: a review.
\newblock {\em Annual Review of Statistics and Its Application}, 4:395--421.

\bibitem[Sampson and Guttorp, 1992]{sampson1992nonparametric}
Sampson, P.~D. and Guttorp, P. (1992).
\newblock Nonparametric estimation of nonstationary spatial covariance
  structure.
\newblock {\em Journal of the American Statistical Association},
  87(417):108--119.

\bibitem[Schmidt et~al., 2011]{schmidt2011considering}
Schmidt, A.~M., Guttorp, P., and O'Hagan, A. (2011).
\newblock Considering covariates in the covariance structure of spatial
  processes.
\newblock {\em Environmetrics}, 22(4):487--500.

\bibitem[Schmidt and O'Hagan, 2003]{schmidt2003bayesian2}
Schmidt, A.~M. and O'Hagan, A. (2003).
\newblock Bayesian inference for non-stationary spatial covariance structure
  via spatial deformations.
\newblock {\em Journal of the Royal Statistical Society: Series B (Statistical
  Methodology)}, 65(3):743--758.

\bibitem[Simpson et~al., 2017]{simpson2017penalising}
Simpson, D., Rue, H., Riebler, A., Martins, T.~G., S{\o}rbye, S.~H., et~al.
  (2017).
\newblock Penalising model component complexity: A principled, practical
  approach to constructing priors.
\newblock {\em Statistical Science}, 32(1):1--28.

\bibitem[Stein, 1999]{ste99}
Stein, M.~L. (1999).
\newblock {\em Statistics for Spatial Data: Some Theory for Kriging}.
\newblock Springer, New York.

\bibitem[Tadi{\'c} et~al., 2015]{tadic2015mapping}
Tadi{\'c}, J.~M., Qiu, X., Yadav, V., and Michalak, A.~M. (2015).
\newblock Mapping of satellite earth observations using moving window block
  kriging.
\newblock {\em Geoscientific Model Development}, 8(10):3311--3319.

\bibitem[Tagle et~al., 2020]{tag20}
Tagle, F., Genton, M., Yip, A., Mostamandi, S., Stenchikov, G., and Castruccio,
  S. (2020).
\newblock A high-resolution bi-level skew-t stochastic generator for assessing
  {S}audi {A}rabia's wind energy resources (with discussion).
\newblock {\em Environmetrics}.
\newblock in press.

\bibitem[Taylor et~al., 2009]{taylor2009wind}
Taylor, J.~W., McSharry, P.~E., and Buizza, R. (2009).
\newblock Wind power density forecasting using ensemble predictions and time
  series models.
\newblock {\em IEEE Transactions on Energy Conversion}, 24(3):775.

\bibitem[Yip, 2018]{yip18}
Yip, C. M.~A. (2018).
\newblock {\em Statistical Characteristics and Mapping of Near-Surface and
  Elevated Wind Resources in the Middle East}.
\newblock PhD thesis, King Abdullah University of Science and Technology.

\bibitem[Zhang, 2004]{zha04}
Zhang, H. (2004).
\newblock Inconsistent estimation and asymptotically equal interpolations in
  model-based geostatistics.
\newblock {\em Journal of the American Statistical Association},
  99(465):250--261.

\end{thebibliography}

\end{document}